\documentclass[headings=standardclasses]{scrartcl}

\usepackage{geometry}
\usepackage[utf8]{inputenc}
\usepackage[T1]{fontenc}
\usepackage[onehalfspacing]{setspace}
\usepackage[charter]{mathdesign}
\usepackage{microtype}
\usepackage{amsmath}
\usepackage{amsthm}
\usepackage{tabulary}
\usepackage{longtable}
\usepackage{booktabs} 
\usepackage{diagbox}
\usepackage{ragged2e}

\usepackage{siunitx}
\usepackage{bm}
\usepackage[pdftex]{graphicx}
\usepackage{rotating}
\usepackage{pdfpages}
\usepackage{pdflscape}
\usepackage{color}
\usepackage{array}
\usepackage[font=bf]{caption}
\usepackage{subcaption}
\usepackage{pgfplots}
\usepackage{pgfplotstable}
\usepackage{float}
\usepackage[title]{appendix}
\pgfplotsset{compat=1.9}
\usepackage{url}
\usepackage[section]{placeins}
\usepackage{natbib}
\usepackage{hyperref}
\hypersetup{
	colorlinks   = true,    % Colours links instead of ugly boxes
	urlcolor     = blue,    % Colour for external hyperlinks
	linkcolor    = blue,    % Colour of internal links
	citecolor    = blue      % Colour of citations
}
\usepackage{algorithm}
\usepackage{algpseudocode}

\makeatletter
\renewcommand*\env@matrix[1][c]{\hskip -\arraycolsep
  \let\@ifnextchar\new@ifnextchar
  \array{*\c@MaxMatrixCols #1}}
\makeatother

\usepackage{xcolor}
\usepackage{soul}

\usepackage{verbatim}

\begin{document}
\thispagestyle{empty}
\begin{spacing}{1.2}
\begin{center}
\huge \textbf{Bridging Models and Modalities: Fusing Models Created from Diverse Multimodal Datasets} \\
\vspace{\baselineskip}
\end{center}
\begin{flushleft}
\normalsize
24 October 2026 \\
\vspace{\baselineskip}
\textsc{Lanlan Yan}  \\
Department of Technology, Management and Economics \\
Technical University of Denmark (DTU), Denmark \\
lanya@dtu.dk \\
\vspace{\baselineskip}
\textsc{Rico Krueger} (corresponding author) \\
Department of Technology, Management and Economics \\
Technical University of Denmark (DTU), Denmark  \\
rickr@dtu.dk \\
\end{flushleft}
\end{spacing}

\newpage
\thispagestyle{empty}
\section*{Abstract}

Transportation choice modelling increasingly involves heterogeneous datasets collected across different experimental settings, raising challenges for combining information to improve predictive performance. These datasets may differ in modality availability, with modality-rich and modality-lean datasets that provide complementary but incomplete views of behaviour. To leverage complementary information across these datasets, we propose a framework for applying Bayesian predictive model-fusion methods to models estimated separately from heterogeneous datasets. We demonstrate this approach using two stated choice datasets on electric versus conventional vehicle choice: a laboratory dataset with eye-tracking measurements and a street-intercept dataset without such measurements. Among the methods considered, Pareto-smoothed importance sampling leave-one-out (PSIS-LOO) stacking achieves the best overall predictive performance. The results further show that the benefits of model fusion depend on the target domain, with the greatest improvements observed for a pooled target domain comprising both experimental settings.
\\
\\
\textbf{Keywords:} model fusion; dataset heterogeneity; shared mobility; discrete choice modelling

\newpage

\section{Introduction}
Predicting individual behaviour in new settings is a central objective of transport choice modelling. Even when models perform well on held-out data, their predictive performance may not necessarily transfer to different contexts. Differences in study populations, experimental settings, and available sources of information may further complicate predictive generalisation.

Transportation choice problems are increasingly investigated across diverse experimental settings, including laboratory experiments, online surveys, virtual reality studies, and field experiments. Across these settings, datasets concern the same type of decision-making problem but may differ in the additional information collected alongside the observed choices. Some experiments collect only observed choices, whereas others additionally include attitudinal indicators \citep[e.g.][]{ton2020latent} or process-related measurements, such as eye-tracking or physiological responses \citep{bogacz2021modelling, argota2024virtual, derakhshan2024situated}. Additionally, differences in recruitment strategies and experimental protocols \citep{lee2022remotelab, mottelson2021self} may result in distinct study populations. Consequently, transport researchers are increasingly faced with heterogeneous datasets that provide different but potentially complementary information about the same choice problem, raising the question of how these sources can be effectively combined to improve predictive performance.

%Methods used in machine-learing cross-domain studies
Related challenges of learning across heterogeneous data sources have received considerable attention in machine learning. For example, heterogeneous domain adaptation and domain generalisation address differences in feature spaces or modalities across domains \citep{rahate2022multimodal}. Existing approaches include cross-dataset and cross-modality transfer learning \citep{ding2015missing,yang2017cross}, generative cross-modal mapping using GANs \citep{athanasiadis2020audio}, and cross-modal representation alignment combined with meta-learning \citep{liang2021cross}. These studies provide relevant adjacent literature, with a primary focus on learning across heterogeneous representations (e.g., images, text, and audio) for tasks such as classification or recognition.

%Methods used in transporation + cross-domain studies
Similar cross-domain learning approaches have also been explored in transportation. For traffic flow prediction, \citet{zhang2025mc} developed a multi-source cross-domain framework to exploit information from multiple source domains, while \citet{shao2025cross} used knowledge distillation and domain adaptation to transfer traffic patterns across different cities. Beyond traffic prediction, \citet{sun2018analyzing} combined semi-supervised and active learning to facilitate knowledge transfer across different transportation data sources. These studies primarily focus on knowledge transfer across domains rather than on differences in modality availability across datasets.

%Methods used in transporation chocie + cross-domain studies
Within transport choice modelling, studies have also examined choice behaviour using datasets collected through different data collection methods or experimental settings. Several approaches have been adopted to analyse or combine information from these datasets. One approach is to analyse the datasets separately, allowing differences in choice behaviour across experimental settings to be examined independently \citep[e.g.][]{burris2023examining}. Another approach is to jointly model multiple datasets, allowing information from different sources to be incorporated into a common choice model, as demonstrated by studies combining stated preference (SP) and revealed preference (RP) data \citep[e.g.][]{ben1990estimation, TABASI2023100632}. Alternatively, information learned from one dataset can be transferred to another, following a strategy similar to the cross-domain transfer methods discussed above. For example, \citet{DING2025104608} transfer attention patterns learned from eye-tracking data collected in the laboratory to a modality-lean street-intercept dataset to improve choice prediction. However, these studies do not explicitly examine the combination of predictions from separately estimated dataset-specific models. 

%Summarize from the heterogeneous datasets in transporation choice modelling
%Taken together, the studies discussed above illustrate different strategies for handling heterogeneous data sources, including separate analysis, joint model estimation, and cross-domain knowledge transfer. While these approaches allow information from different datasets to be examined or integrated in different ways, they do not directly address the possibility of retaining dataset-specific models and combining the information they provide at the prediction level. 

%suggests prediction-level fusion as another strategy
Prediction-level model fusion provides an alternative strategy by combining predictions from separately estimated models. Its potential has recently been demonstrated in transport choice forecasting by \citet{hess2026combine}, who use model averaging to improve predictive performance, particularly for out-of-distribution predictions. For heterogeneous datasets, this strategy allows dataset-specific models to retain specifications appropriate to their available information, without requiring joint training or a shared feature space.

%A methodological framework that we can use for predictive-level model fusion
Models estimated from heterogeneous datasets may differ in their parameter uncertainty due to differences in sample size, available information, and model specification. Bayesian predictive modelling provides a convenient basis for model fusion by incorporating this uncertainty into each model's posterior predictive distribution before combining predictions. For fusing Bayesian predictive distributions, \citet{yao2018using} examine several approaches, including stacking, pseudo-Bayesian model averaging (pseudo-BMA), and Bayesian-bootstrap pseudo-BMA (BB-pseudo-BMA). These methods weigh candidate models according to their predictive performance, with stacking selecting weights to maximize the expected predictive accuracy of the combined predictive distribution.

%predictive model fusion are used for one common dataset ---> gap: heteronegous datasets
However, both the transport choice application of \citet{hess2026combine} and the conventional settings considered by \citet{yao2018using} involve candidate models estimated and evaluated using a common dataset. Our setting instead requires fusing models learned from heterogeneous datasets describing the same choice problem. Consequently, there is no single dataset that naturally defines the domain over which their predictive performance should be evaluated and combined. We therefore explicitly define the \textbf{target domain} as the population and experimental setting(s) for which predictive performance is to be optimised. This provides a common basis for evaluating and combining models trained on different datasets. To the best of our knowledge, predictive model fusion has not been explicitly considered for models learned from modality-rich and modality-lean datasets describing the same decision problem.

%Research aim
Accordingly, the central research aim of this study is to develop a framework that applies existing predictive model-fusion methods to models estimated from modality-rich and modality-lean datasets describing the same choice problem. We consider the methods from \citet{yao2018using}, with particular emphasis on Pareto-smoothed importance sampling leave-one-out (PSIS-LOO) stacking. The proposed framework estimates model-fusion weights for a specified target domain and evaluates the resulting fused predictive distribution on held-out data from that domain. Although demonstrated in transport choice modelling, this approach may also be applicable to other prediction problems involving heterogeneous datasets with a common outcome, such as consumer choice modelling using online surveys and laboratory experiments with eye-tracking measurements. 

The remainder of this paper is structured as follows. Section~\ref{sec:method} presents the model-fusion methodology, followed by the empirical application in Section~\ref{sec:application}. Section~\ref{sec:results} presents the results and discussion, and Section~\ref{sec:conclusion} concludes the paper. Additional implementation details and sensitivity analyses are provided in the appendices.

\section{Methodology}
\label{sec:method}
This section first introduces the three predictive model-combination methods considered in this study: Pareto-smoothed importance sampling leave-one-out (PSIS-LOO) stacking, pseudo-Bayesian model averaging (pseudo-BMA), and Bayesian-bootstrap pseudo-BMA (BB-pseudo-BMA) \citep{yao2018using}, with PSIS-LOO stacking as our primary method. We then present the proposed model-fusion framework, which extends these methods to models estimated from heterogeneous datasets by formulating model fusion with respect to a defined target domain.

\subsection{Bayesian predictive model fusion methods}
\begin{figure}[htbp]
    \centering
    \includegraphics[width=0.8\textwidth]{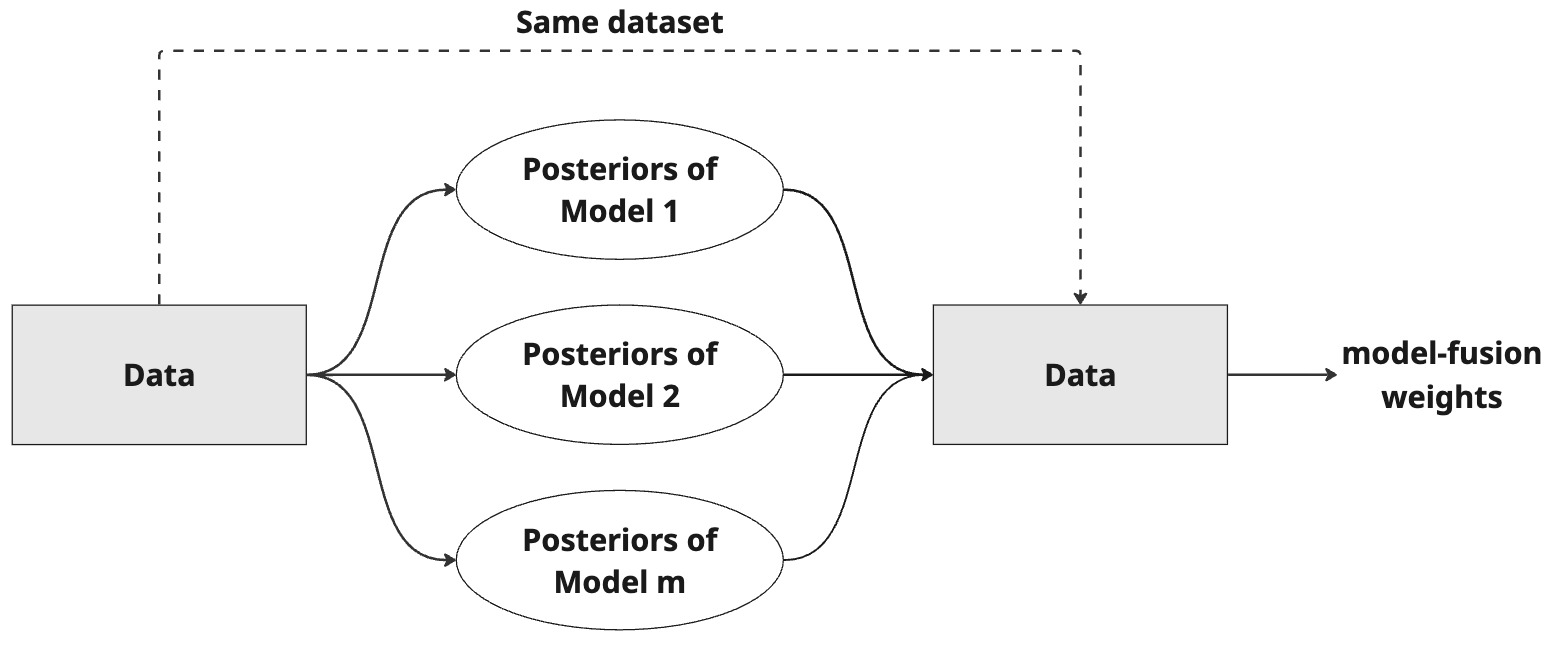}
    \caption{Conventional single-dataset workflow for predictive model fusion, from \cite{yao2018using}}
    \label{fig:flow_original}
\end{figure}
Figure \ref{fig:flow_original} illustrates the conventional single-dataset setting of the model-fusion methods considered in this study. A common dataset is used to estimate the posterior distributions of the candidate models and, through leave-one-out (LOO) prediction, to evaluate their predictive performance for learning the model-fusion weights. Specifically, the model posteriors are used to construct pointwise LOO predictive densities, which form the basis for weight estimation. The three methods differ in how these predictive densities are used to determine the model-fusion weights, as described below.

Suppose that $M$ candidate models are available, each with predictive density $p_m(y_i \mid x_i)$ for observation $i$, where $m=1,\dots,M$. Let
\begin{equation}
\ell_{im} = \log p_m(y_i \mid y_{-i}, x_i)
\end{equation}
denote the leave-one-out (LOO) log predictive density, approximated using Pareto-smoothed importance sampling (PSIS). The corresponding expected log predictive density (ELPD) is
\begin{equation}
\widehat{\mathrm{ELPD}}_m = \sum_{i=1}^{N} \ell_{im}.
\end{equation}

PSIS-LOO stacking combines predictive distributions through a weighted mixture
\begin{equation}
p(y \mid x)
=
\sum_{m=1}^{M} w_m \, p_m(y \mid x),
\quad
w_m \geq 0,\ \sum_{m=1}^{M} w_m = 1,
\end{equation}
where the weights are obtained by maximizing the LOO predictive performance of the combined distribution:
\begin{equation}
\hat{w}
=
\arg\max_{w}
\sum_{i=1}^{N}
\log\left(
\sum_{m=1}^{M} w_m \exp(\ell_{im})
\right).
\end{equation}

In contrast, pseudo-BMA assigns weights based on model-level ELPD:
\begin{equation}
w_m^{\mathrm{pBMA}}
=
\frac{\exp(\widehat{\mathrm{ELPD}}_m)}
{\sum_{j=1}^{M}\exp(\widehat{\mathrm{ELPD}}_j)}.
\end{equation}
BB-pseudo-BMA extends this approach by incorporating uncertainty in ELPD via Bayesian bootstrap.

The key distinction is that stacking operates at the observation level by directly optimizing the predictive performance of the mixture distribution, whereas pseudo-BMA and BB-pseudo-BMA rely on aggregated model-level performance. This gives stacking greater flexibility to exploit complementary predictive patterns across models, particularly when different models perform well for different observations.

\subsection{Model fusion framework for heterogeneous datasets}
\begin{figure}[H]
    \centering
    \includegraphics[width=1\textwidth]{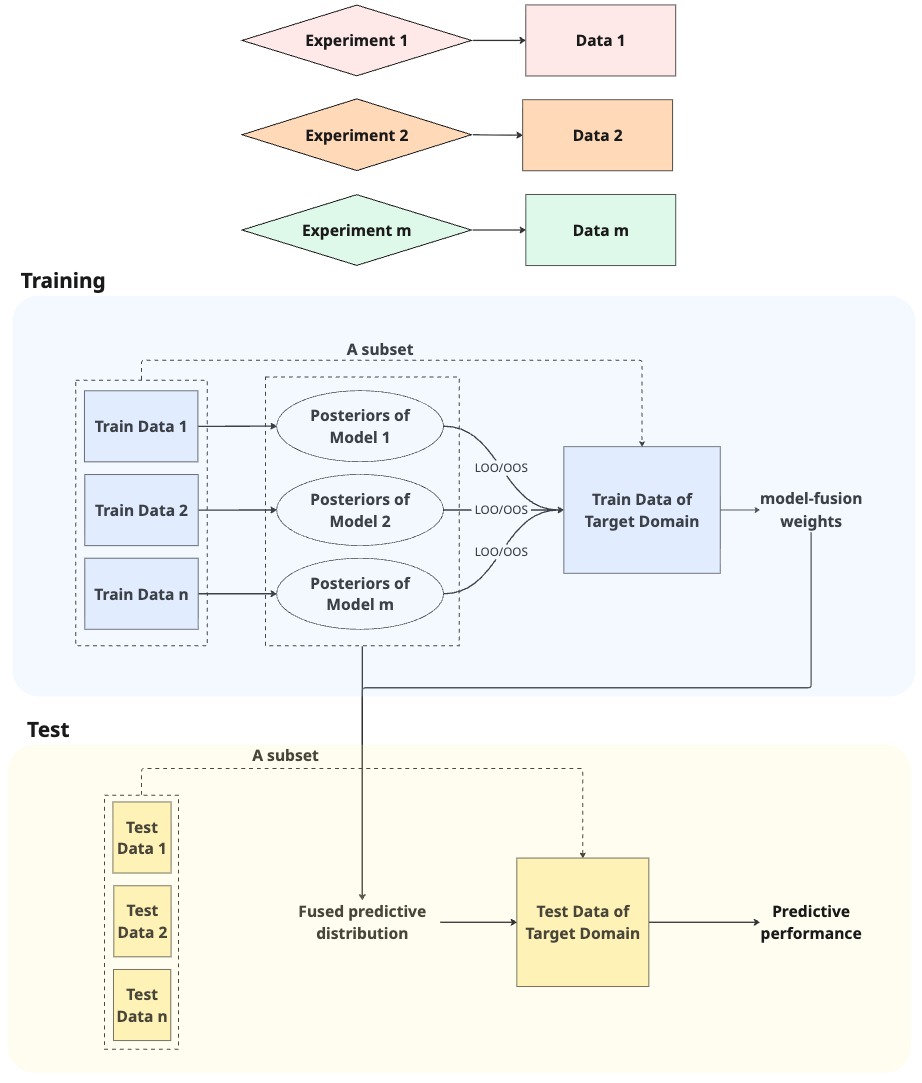}
    \caption{Model-fusion framework for multiple heterogeneous datasets}
    \label{fig:flow_general}
\end{figure}

The proposed framework extends conventional model-fusion methods in two main respects. First, it introduces an explicit separation between model-fusion weight estimation and predictive evaluation. Each dataset is divided into training and test sets: candidate models and model-fusion weights are learned using the training data, while the resulting fused predictions are evaluated on held-out test data. Second, candidate models are estimated from separate heterogeneous datasets describing the same underlying prediction problem, rather than from a common dataset. Model fusion is therefore formulated with respect to a \textbf{target domain}, defined as the population and experimental setting(s) for which predictive performance is to be optimised. The target domain may correspond to a single experimental domain or a pooled domain comprising observations from multiple experimental settings. Different target domains define different domains over which predictive performance is optimised and may therefore lead to different model fusion weights. Figure \ref{fig:flow_general} illustrates the general framework.

Within this framework, the behavioural model is specified and estimated separately for each training dataset under a Bayesian framework, yielding model-specific posterior distributions. For a given target domain, pointwise predictive densities are constructed for each model using (1) leave-one-out (LOO) estimates for observations from the model's own estimation dataset and/or (2) out-of-sample (OOS) predictions for observations from other datasets. The latter are required because some target-domain observations were not used to estimate a given candidate model and therefore cannot be evaluated through LOO prediction, representing a practical form of out-of-distribution (OOD) prediction when the datasets differ in experimental settings. Based on these predictive densities, model-fusion weights are computed using PSIS-LOO stacking, pseudo-BMA, and BB-pseudo-BMA. Finally, the learned model-fusion weights are combined with the model posterior predictive distributions to form the fused predictive distribution, which is evaluated on the held-out test data of the target domain to obtain its predictive performance. 

An additional consideration arises when datasets collected under different experimental settings differ in their available modalities (e.g., process data versus choice-only data).
Some models may require inputs that are unavailable in other datasets. An alignment mechanism is therefore required to enable these models to generate comparable predictions across datasets. The specific form of this alignment may vary depending on the application. Section \ref{sec:application} describes its implementation in the present study. 

\section{Application}
\label{sec:application}
This section applies the proposed model-fusion framework to two stated choice datasets collected by \citet{DING2025104608} under two different experimental settings: i) a modality-rich laboratory dataset with eye-tracking information and ii) a modality-lean street-intercept dataset containing choice data only. We first describe the two datasets and examine the visual-attention information available in the laboratory data. Then, we specify the dataset-specific behavioural models and show how this difference in available modalities creates a structural mismatch when applying the proposed framework. An attention-transfer mechanism is introduced to align the model inputs across datasets, after which we define the two target domains used for model-fusion weight estimation and predictive evaluation.
\subsection{Dataset}
%General description of the dataset%
The data consist of stated choice observations collected under two experimental settings. A total of 152 ride-hailing drivers participated in street-intercept experiments, while 40 drivers participated in a laboratory experiment with eye-tracking measurements. Participants completed 2 (street) and 5 (lab) choice tasks respectively, resulting in 504 observations in total. Each task involves a choice between a conventional vehicle (CV) and an electric vehicle (EV) described by attributes including rental cost (RC), operating cost (OC, uncertain for EV), and driving range (DR). The two samples exhibit similar demographic characteristics, suggesting a relatively consistent population across datasets. Table \ref{tab:dataset_overview} summarises the main characteristics of the two datasets. 
\begin{table}[htbp]
    \centering
    \caption{Overview of the lab and street-intercept stated-choice datasets.}
    \label{tab:dataset_overview}
    \begin{tabular}{lcc}
        \hline
        \textbf{Characteristic} & \textbf{Lab dataset} & \textbf{Online dataset} \\
        \hline
        Participants              & 40  & 152 \\
        Choice tasks              & 200 & 304 \\
        Tasks per participant     & 5   & 2 \\
        Choice data               & Yes & Yes \\
        Eye-tracking data         & Yes & No \\
        \hline
    \end{tabular}
\end{table}

%Eye-tracking data and attention-share construction%
The principal modality difference between the two datasets is the availability of visual-attention information in the laboratory experiment. We therefore briefly describe its representation before specifying the behavioural models. Visual attention in the lab experiment was measured using fixation counts for the four attributes considered in the choice model: RC, OC1, OC2, and DR. Specifically, the attention share for attribute $k$ in choice task $t$ is defined as
\begin{equation}
    A_{kt}
    =
    \frac{N_{kt}}
    {\sum_{j=1}^{K} N_{jt}},
    \qquad K=4,
    \label{eq:attention_share}
\end{equation}
where $N_{kt}$ denotes the number of fixations on attribute $k$ in task $t$. Note that when the total fixation count is zero, equal attention shares of 0.25 are assigned to all four attributes.

\begin{figure}[H]
    \centering
    \includegraphics[width=0.8\textwidth]{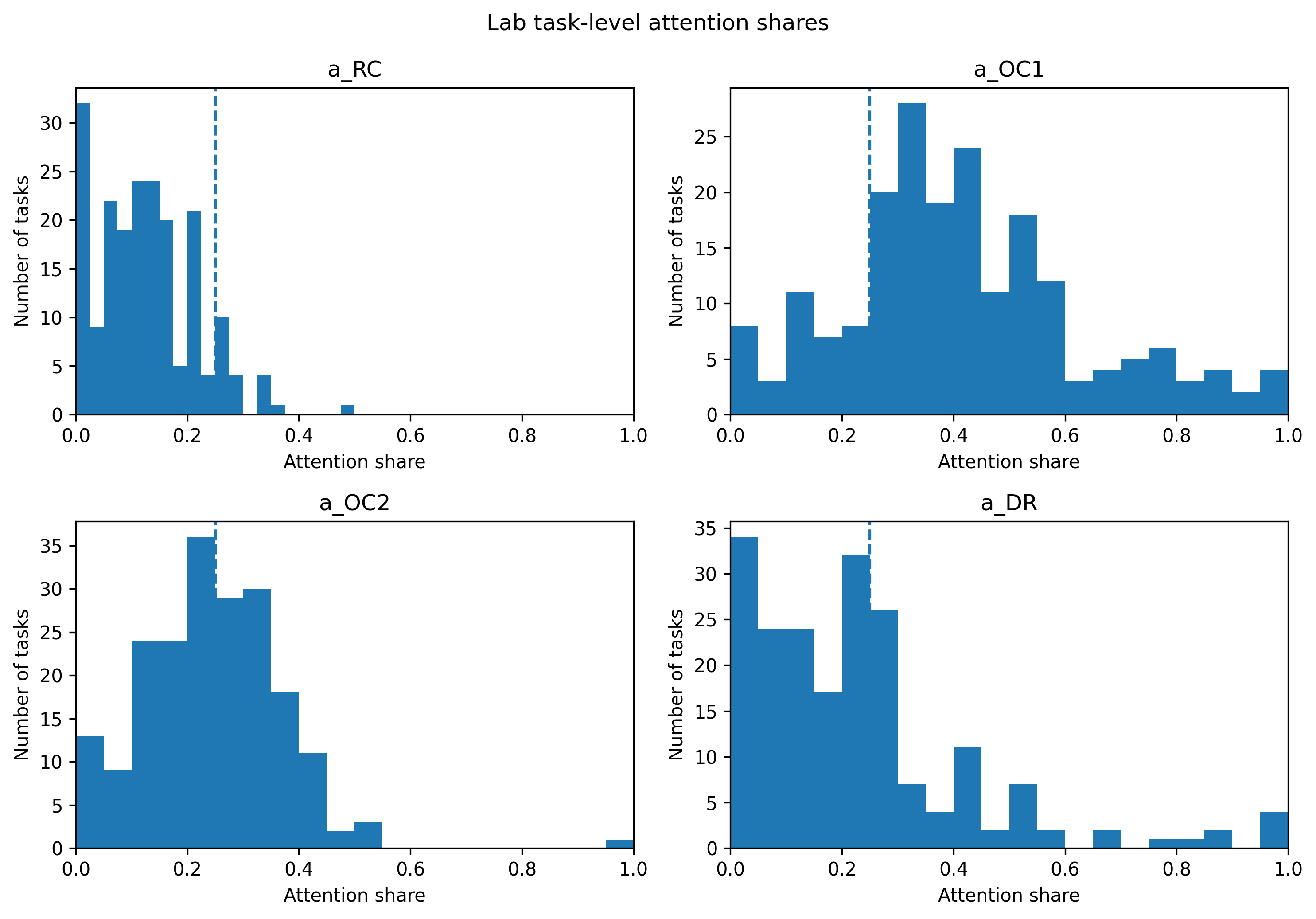}
    \caption{Distribution of task-level visual-attention shares across the four attributes in the laboratory dataset. Dashed lines indicate equal attention allocation (\(A_{kt}=0.25\))}
    \label{fig:attention_distribution}
\end{figure}

%Attention distribution figure + 0.25 justification%
Figure~\ref{fig:attention_distribution} shows the distributions of the task-level attention shares. The average shares were 0.124 for RC, 0.411 for OC1, 0.240 for OC2, and 0.224 for DR, indicating substantial differences in attention allocation across attributes. Since the four shares sum to one within each task, \(0.25\) represents equal attention allocation and is used as a common reference point. We therefore define the attention deviation for attribute $k$ in task $t$ as
\begin{equation}
    \widetilde{A}_{kt}
    =
    A_{kt}-0.25.
    \label{eq:centered_attention}
\end{equation}

Thus, $\widetilde{A}_{kt}=0$ represents equal attention allocation, while positive and negative values indicate above- and below-equal attention, respectively. This transformation is used in the laboratory MCPT specification introduced below.

\subsection{Base behavioural model and structural mismatch}
We next specify the two dataset-specific behavioural models for model fusion. Both are based on the multi-attribute cumulative prospect theory (MCPT) model \citep{avineri2004cumulative,de2011expected,tversky1992advances,xu2011decision}, which has also been adopted by \citet{DING2025104608}, but differ in whether visual attention enters the attribute sensitivities. This difference reflects the information available in the respective datasets and gives rise to the structural mismatch considered below.

\paragraph{Street MCPT model (7 parameters)}
For the street MCPT model, let $\mathbf{x}_i$ denote the observed attribute vector for choice task $i$, and let $\boldsymbol{\theta}_{\mathrm{street}}$ denote the corresponding model parameters. For compactness, the street MCPT utility of alternative $j$ is represented as 
\begin{equation}
U_{ij}
=
ASC_j
+
\sum_{k}
v\!\left(
\beta_{ik}
(x_{ijk}-r_{ik})
\right),
\label{eq:utility}
\end{equation}
where $v(\cdot)$ denotes the reference-dependent MCPT prospect value transformation, and uncertain attribute outcomes are incorporated through Prelec probability weighting. For the street model, the attribute sensitivities are assumed to be constant,
\begin{equation}
\beta_{ik}=\beta_k,
\label{eq:web_beta}
\end{equation}
leading to the predictive distribution
\begin{equation}
p_{\mathrm{street}}
\left(
y_i
\mid
\mathbf{x}_i,
\boldsymbol{\theta}_{\mathrm{street}}
\right).
\label{eq:web_predictive}
\end{equation}

The complete MCPT formulation, including the prospect value function, probability weighting, reference points, and prior specifications, is provided in \ref{app:model_specification}.

\paragraph{Laboratory MCPT model (10 parameters)}
The laboratory model shares the same MCPT utility specification as Eq.~(\ref{eq:utility}) but replaces the constant attribute sensitivities with attention-dependent coefficients,
\begin{equation}
\beta_{ik}
=
\operatorname{softplus}
\left(
\beta_{k,0}
+
\gamma_k
(A_{ik}-0.25)
\right),
\label{eq:attention_beta}
\end{equation}
where $A_i$ denotes the task-level attention vector, $\beta_{k,0}$ is the baseline component of attribute sensitivity, and $\gamma_k$ captures the influence of attention on attribute sensitivity. Attention is centred at 0.25, corresponding to an equal allocation of attention across the four attribute-level attention measures. The corresponding predictive distribution is therefore
\begin{equation}
p_{\mathrm{lab}}
\left(
y_i
\mid
\mathbf{x}_i,
A_i,
\boldsymbol{\theta}_{\mathrm{lab}}
\right),
\label{eq:lab_predictive}
\end{equation}
which differs from the street model only through the specification of the attribute sensitivities. The complete model specification, including parameter definitions and prior distributions, is provided in \ref{appendixA}.

\paragraph{Structural mismatch}
A key challenge is that attention information is only observed in the laboratory dataset. As a result, the predictive distributions differ in their information sets:
\begin{equation}
p_{\text{lab}}(y_i \mid x_i, A_i, \theta_{\text{lab}}),
\qquad
p_{\text{street}}(y_i \mid x_i, \theta_{\text{street}}).
\end{equation}
This discrepancy creates a structural mismatch: the two models are defined on different inputs, making it difficult to apply the lab model to street data without further adjustment.

\subsection{Attention transfer and model alignment}

To address this issue, we introduce an attention-transfer mechanism as a structural alignment step, rather than an attention estimation procedure. We assume that scenario structure provides a reasonable approximation of average attention allocation. Accordingly, for each scenario $s$, we compute the average attention vector from the laboratory data across observations sharing the same scenario, where attention weights are derived from fixation counts:
\begin{equation}
\bar{A}_s = \mathbb{E}(A_i \mid s_i = s, D_{\text{lab}}).
\end{equation}
For each street observation, the unobserved attention is approximated by assigning the corresponding scenario-level average, i.e.,
$A_i^{\text{street}} \approx \bar{A}_{s_i}$, yielding an augmented predictive model
\begin{equation}
p_{\text{lab}}(y_i \mid x_i, \bar{A}_{s_i}, {\theta}_{\mathrm{lab}}),
\end{equation}
which is structurally aligned with the laboratory-based model.

The stacked predictive distribution can be written as
\begin{equation}
p(y_i \mid x_i) =
w_{\text{street}}\, p_{\text{street}}(y_i)
+
w_{\text{lab}}\, p_{\text{lab}}(y_i),
\end{equation}
Here, \(p_{\mathrm{street}}(y_i)\) and \(p_{\mathrm{lab}}(y_i)\) denote the predictive densities for the same observation under the street and laboratory models, respectively. The street model uses only the choice attributes, whereas the laboratory model uses observed attention for laboratory observations and transferred scenario-level attention for street observations.

Moreover, the proposed attention-transfer mechanism is evaluated through a sensitivity analysis presented in \ref{sensitivity_analysis_attention}, where several alternative transfer strategies are considered. The results show that the predictive performance is largely insensitive to the specific transfer strategy, supporting the assumption of attention transfer as a solution to structural mismatch. 

The attention transfer mechanism is applied uniformly across the target domains of this study and may serve as a general solution to reconcile modality differences in other cases.

\subsection{Target domains and evaluation settings}

\begin{figure}[H]
    \centering
    \includegraphics[width=1\textwidth]{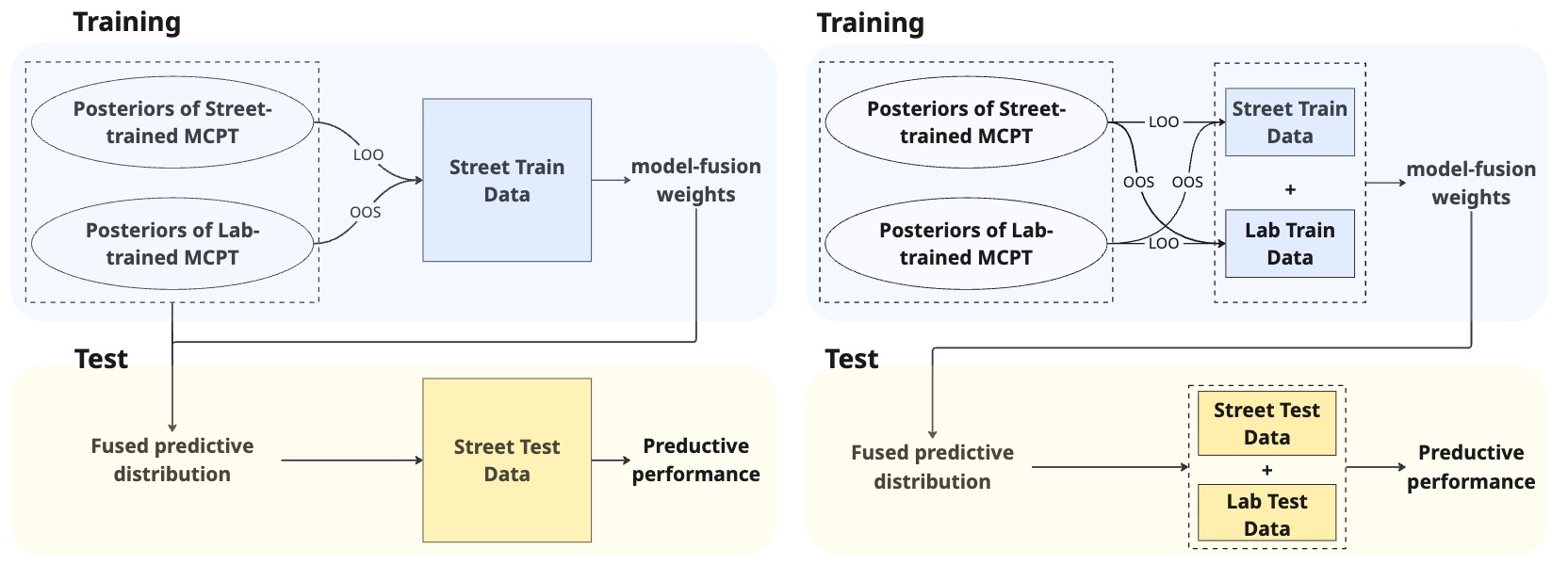}
    \caption{Target domains and evaluation settings for the application to the datasets from \cite{DING2025104608}}
    \label{fig:flow}
\end{figure}

Figure \ref{fig:flow} illustrates the two target domains considered in this application, which define the data used for model-fusion weight estimation and predictive evaluation. For both domains, the laboratory and street-intercept datasets are each split by individual into 80\% training and 20\% test sets.

We consider two target domains:

\textbf{Target Domain 1 (street-only):} This domain includes only street-intercept observations (152 participants $\times$ 2 choice tasks = 304 observations) and represents a setting where performance is evaluated on modality-lean data. It allows us to assess whether information from the modality-rich (but smaller) laboratory dataset improves prediction for street data.

\textbf{Target Domain 2 (pooled):} This domain includes both laboratory (40 participants $\times$ 5 choice tasks = 200 observations) and street-intercept observations (152 participants $\times$ 2 choice tasks = 304 observations), corresponding to approximately 40\% laboratory and 60\% street-intercept observations. It represents a pooled setting where performance is evaluated over the combined data distribution. It allows us to assess whether modality-rich and modality-lean models provide complementary predictive value across experimental settings.

For weight estimation in Target Domain 1, predictions from Street-trained MCPT are obtained using LOO, while Lab-trained MCPT is evaluated OOS on the street training data. In Target Domain 2, LOO predictions are used for observations from each model's own estimation dataset, while OOS predictions are used for observations from the other dataset.

\section{Results and discussion}
\label{sec:results}
This section is organised into five parts. First, we compare the three model fusion methods with the baseline models. Second, we examine performance differences between Target Domain 1 and Target Domain 2. Third, we analyse the contributions of the two base models to the stacked predictions. Fourth, we implement a sensitivity analysis on the composition of street and lab observations in the pooled target domain. Finally, we extend stacking to a logit-based formulation with individual-level weights and compare it with standard stacking.

\begin{table}[ht]
\centering
\caption{Predictive performance and stacking weights (Target Domain 1: Street data only)}
\label{tab:domain1}
\begin{tabular}{p{4cm}cccc}
\hline
Model & Avg. Log Score & Variance & Weights\_Street &Weights\_Lab \\
\hline
Street-MCPT (baseline, MLE)  & -0.4936 & 0.5863 & -- & -- \\
Pooled-MCPT (street + lab, baseline, MLE) & -0.5452 & 0.1159 & -- & -- \\
Pseudo-BMA & -0.4666 & 0.2101 & 1.0000 & 0.0000 \\
BB-pseudo-BMA & -0.4666 & 0.2101 & 1.0000 & 0.0000 \\
Stacking & \textbf{-0.4666} & 0.2101 & 1.0000 & 0.0000 \\
\hline
\end{tabular}
\end{table}

\begin{table}[ht]
\centering
\caption{Predictive performance and stacking weights (Target Domain 2: Pooled Street and Lab data)}
\label{tab:domain2}
\begin{tabular}{p{4cm}cccc}
\hline
Model & Avg. Log Score & Variance & Weights\_Street &Weights\_Lab \\
\hline
Street-MCPT (baseline, MLE) & -0.7882 & 1.2200 & -- & -- \\
Lab-MCPT (baseline, MLE) & -0.7981 & 0.1032 & -- & -- \\
Pooled-MCPT (street + lab, baseline, MLE) & -0.6661 & 0.2624 & -- & -- \\
Pseudo-BMA & -0.6249 & 0.4069 & 1.0000 & 0.0000 \\
BB-pseudo-BMA & -0.6172 & 0.3391 & 0.9405 & 0.0595 \\
Stacking & \textbf{-0.6132} & 0.1502 & 0.6899 & 0.3101 \\
\hline
\end{tabular}
\end{table}

\begin{figure}[htbp]
    \centering
    
    \begin{subfigure}[t]{0.48\textwidth}
        \centering
        \includegraphics[width=\textwidth]{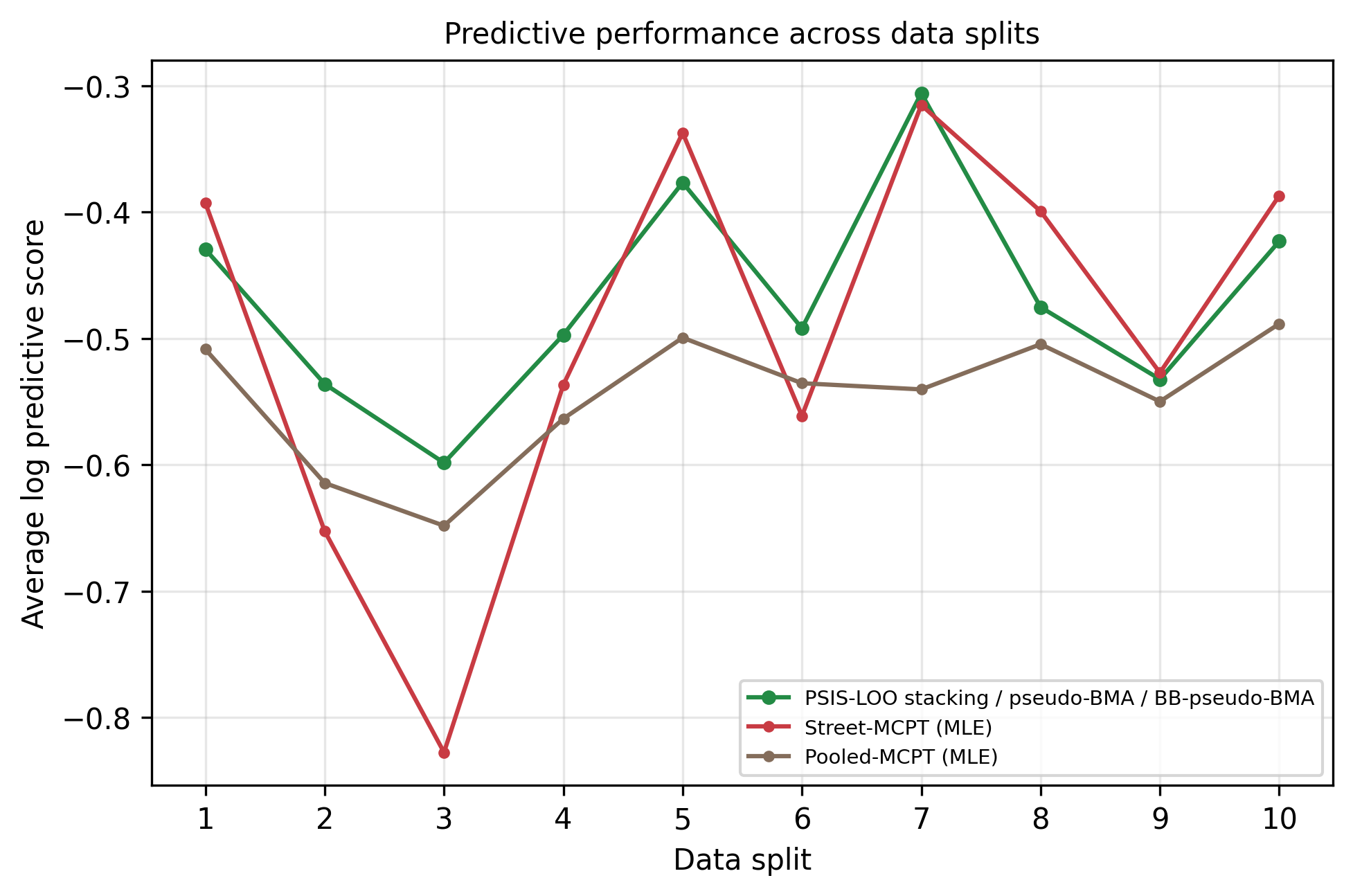}
        \caption{Predictive performance across data splits}
        \label{fig:domain1_splits}
    \end{subfigure}
    \hfill
    \begin{subfigure}[t]{0.48\textwidth}
        \centering
        \includegraphics[width=\textwidth]{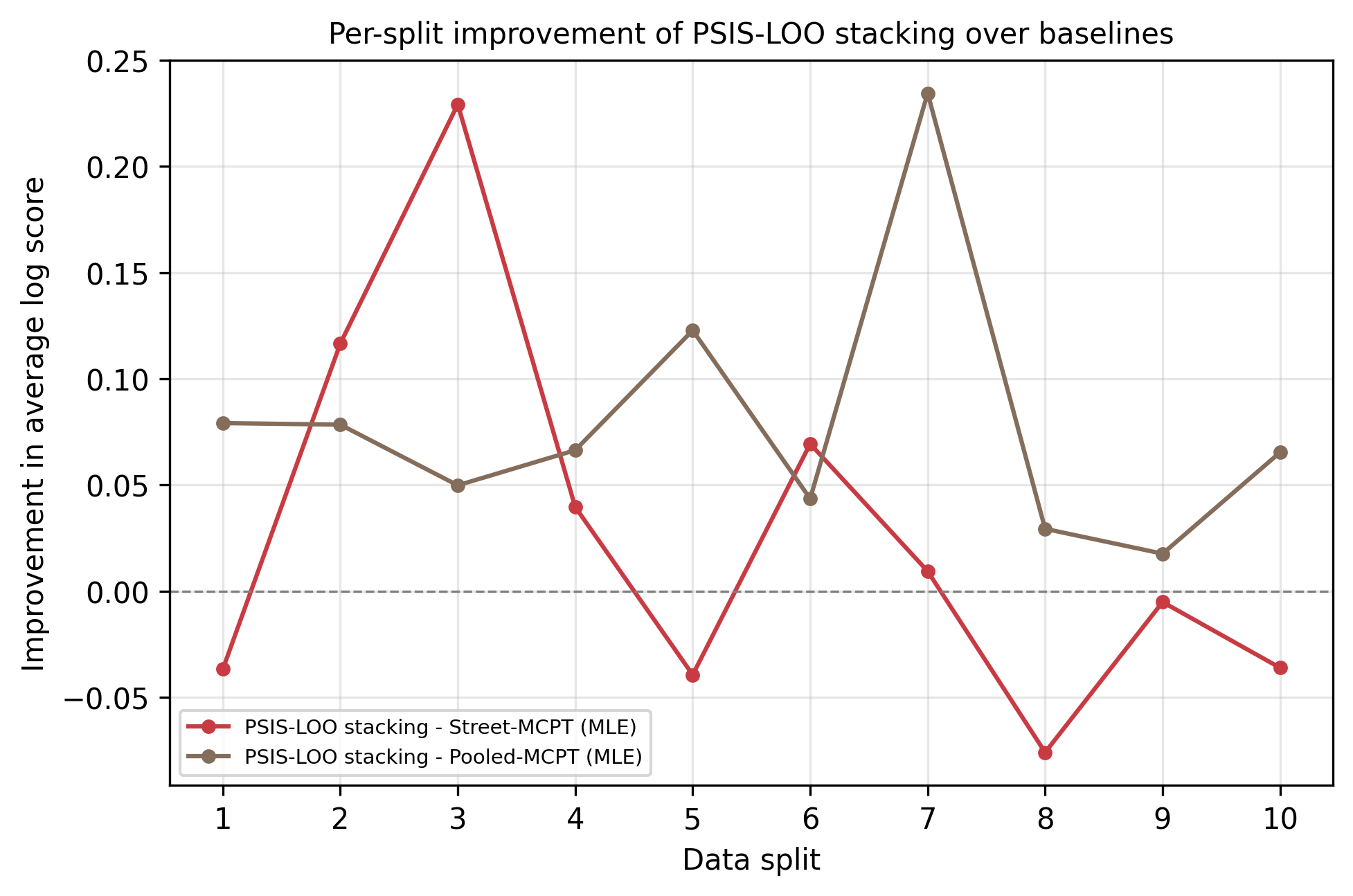}
        \caption{Per-split improvement over baseline models}
        \label{fig:domain1_improvement}
    \end{subfigure}
    
    \caption{Predictive performance comparison in Domain 1}
    
    \label{fig:domain1_results}
\end{figure}

\begin{figure}[htbp]
    \centering
    
    \begin{subfigure}[t]{0.48\textwidth}
        \centering
        \includegraphics[width=\textwidth]{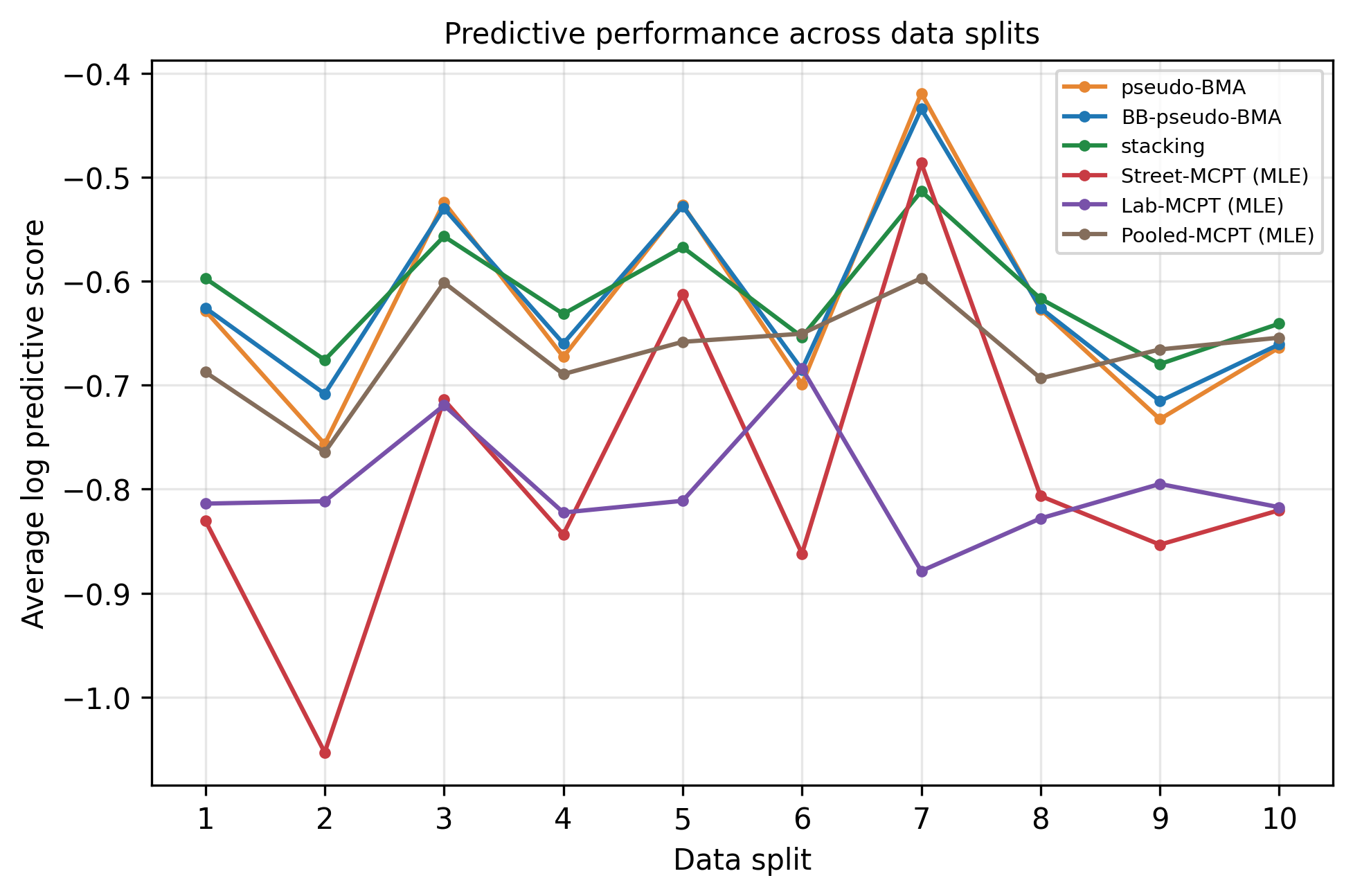}
        \caption{Predictive performance across data splits}
        \label{fig:domain2_splits}
    \end{subfigure}
    \hfill
    \begin{subfigure}[t]{0.48\textwidth}
        \centering
        \includegraphics[width=\textwidth]{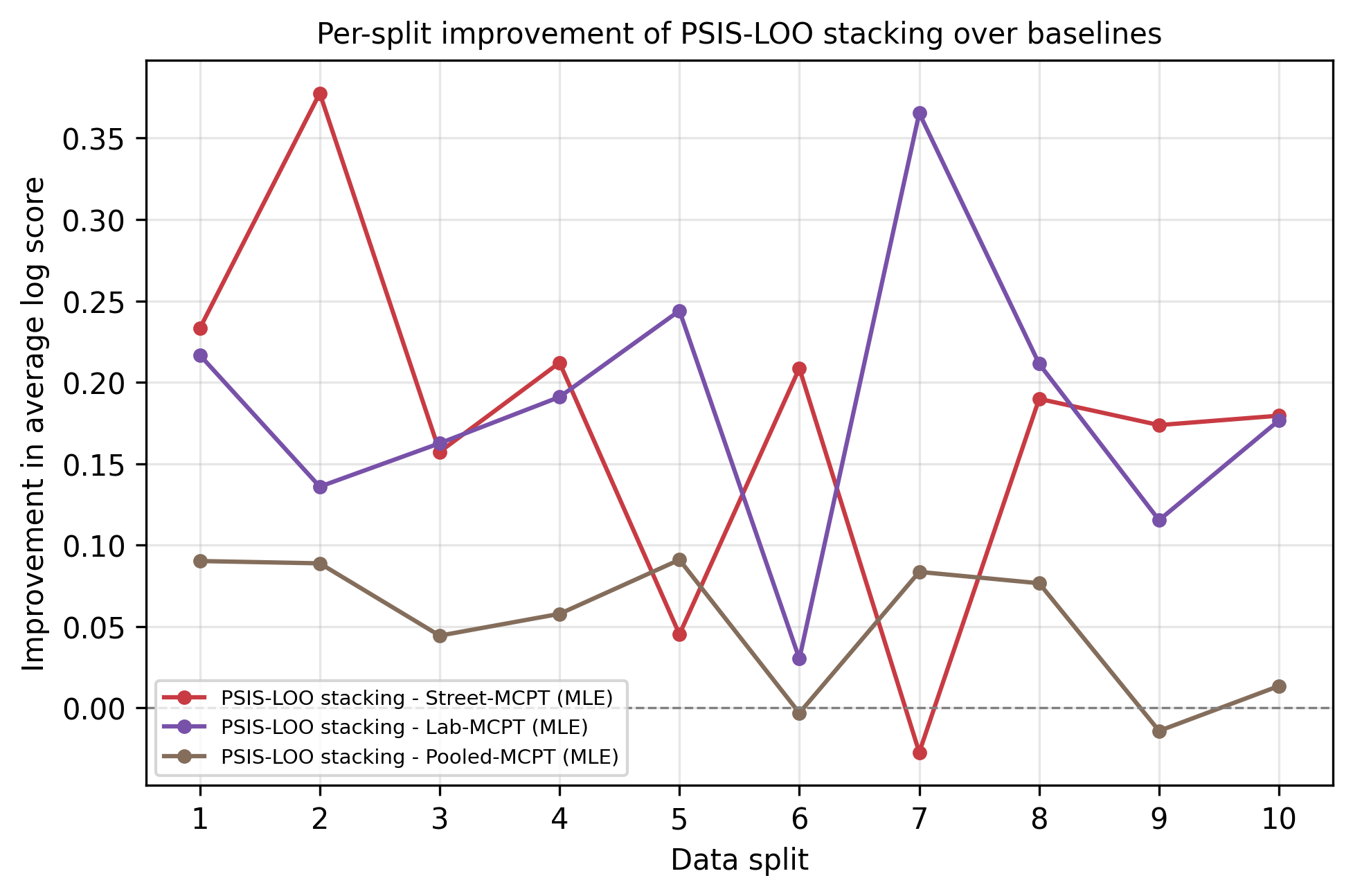}
        \caption{Per-split improvement over baseline models}
        \label{fig:domain2_improvement}
    \end{subfigure}
    
    \caption{Predictive performance comparison in Domain 2}
    
    \label{fig:domain2_results}
\end{figure}

\subsection{Baseline models vs. model fusion}
\label{subsec:baseline_models_vs_model_fusion}
To evaluate whether model fusion improves predictive performance, we compare three fusion methods with several baseline MCPT models in both target domains (Tables \ref{tab:domain1} and \ref{tab:domain2}). Results are averaged over 10 data splits. The baseline models differ in their training data. In Target Domain 1, both models are evaluated on street-intercept test data: \textit{Street-MCPT} is trained on street data only, while \textit{Pooled-MCPT} is trained on the combined dataset, comprising 160 (40$\times$0.8$\times$5) laboratory and 244 (152$\times$0.8$\times$2) street-intercept observations (approximately 40\% and 60\%, respectively). In Target Domain 2, all models are evaluated on the pooled test set, including \textit{Street-MCPT}, \textit{Lab-MCPT}, and \textit{Pooled-MCPT}. All baseline models are estimated using MLE. 

Across both target domains, fusion methods consistently outperform all baseline models in terms of average log predictive score. In Target Domain 1, \textit{Street-MCPT} outperforms \textit{Pooled-MCPT}, indicating that adding lab data may introduce noise when predicting street-based observations. In Target Domain 2, \textit{Pooled-MCPT} performs better than single-source models, as the test data includes both lab and street observations.

Figures \ref{fig:domain1_splits} and~\ref{fig:domain2_splits} show that stacking generally achieves competitive predictive performance across data splits, while baseline models exhibit higher variability. Although stacking does not outperform every baseline in every data split, it achieves higher average predictive performance across the ten splits in both target domains. Among the three fusion methods, PSIS-LOO stacking performs best overall, in line with \cite{yao2018using}.

\subsection{Target Domain 1 vs. Target Domain 2}
In Target Domain 1, all fusion methods yield identical results and assign full weight to the street-based model (Table \ref{tab:domain1}). This indicates that the lab-based model does not provide additional predictive value when the target domain consists only of street-intercept data. As a result, model fusion collapses to the street model.

In Target Domain 2, the fusion methods produce different weights and performances (Table~\ref{tab:domain2}). In particular, stacking assigns weights of 0.690 to the street-based model and 0.310 to the lab-based model. These weights differ from the composition of the pooled target domain, which contains approximately 60\% street and 40\% laboratory observations, indicating that the stacking weights are not simply proportional to the dataset shares but reflect the relative predictive contributions of the two models. The non-zero weight assigned to the lab-based model suggests that it provides complementary predictive information in the pooled domain. In contrast, pseudo-BMA and BB-pseudo-BMA tend to favour the street-based model more strongly.

From a performance perspective, Figures \ref{fig:domain1_improvement} and \ref{fig:domain2_improvement}  show that stacking achieves more consistent improvements over the MLE baselines across data splits in Target Domain 2. This may suggest that model fusion is more beneficial in a pooled target domain.

Overall, these results suggest that the effectiveness of model fusion depends on the target domain. When the domain is homogeneous, fusion may reduce to a single model, whereas in heterogeneous settings, stacking can effectively combine complementary information from different data sources. 

\subsection{Understanding the gains from model stacking}
\label{subsec:4.3}
Among the three fusion methods, PSIS-LOO stacking achieves the highest average predictive performance in Target Domain 2, while all three methods perform equally in Target Domain 1. We therefore \textbf{focus on stacking in Target Domain 2} for the following analysis to understand how it combines the predictive distributions of the two Bayesian candidate models. Unlike Target Domain 1, where stacking effectively reduces to the street-based model (Table \ref{tab:domain1}), Target Domain 2 provides a more informative setting for examining model fusion across different experimental settings.

\begin{figure}[H]
    \centering
    \includegraphics[width=0.8\textwidth]{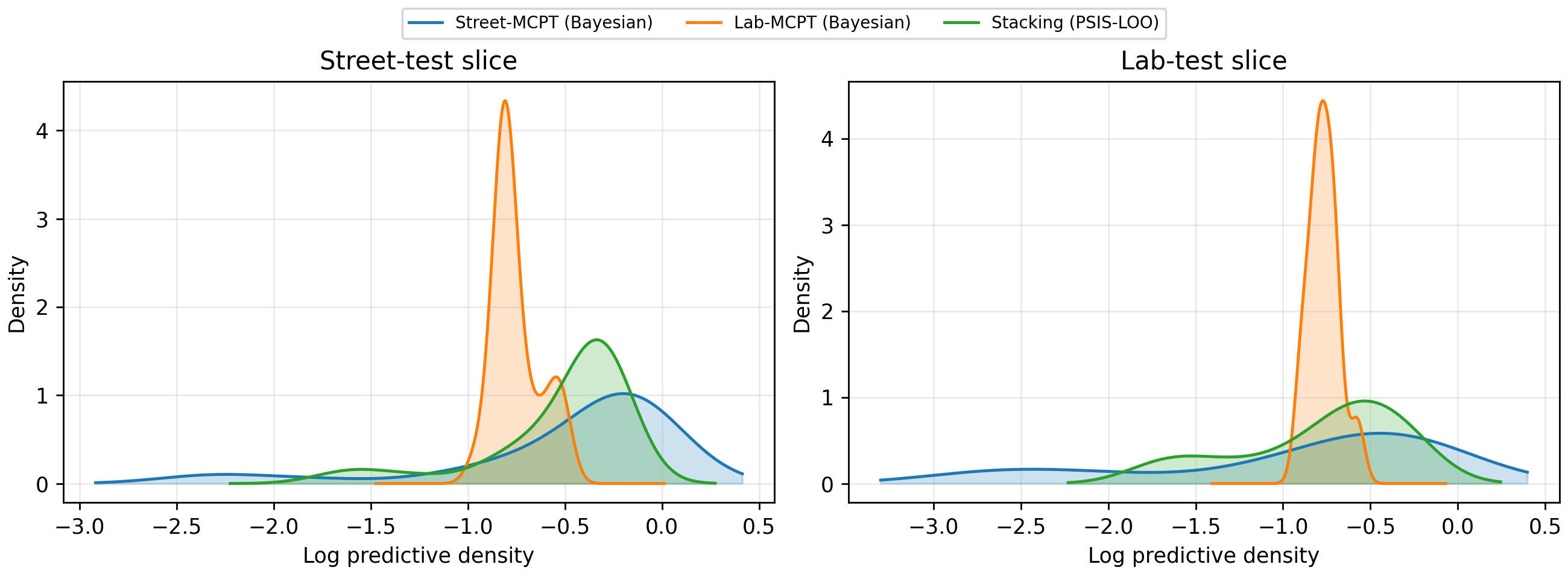}
    \caption{Distribution of pointwise log predictive densities in Target Domain 2 (a representative seed)}
    \label{fig:stacking_distribution}
\end{figure}

Table \ref{tab:domain222} reports average log predictive scores in Target Domain 2 (over 10 splits). The pooled score is calculated as the average log predictive score across all test observations, combining the street-test and lab-test subsets according to their respective sample sizes. Figure \ref{fig:stacking_distribution} shows the distribution of pointwise log predictive densities for a representative split. Unlike the MLE baselines used for benchmarking in Section \ref{subsec:baseline_models_vs_model_fusion}, these Bayesian candidate models are the components directly combined through stacking. They jointly illustrate the predictive patterns in Target Domain 2. The \textit{Street-MCPT (Bayesian)} model performs well on street-test data but deteriorates substantially on lab-test data, while the \textit{Lab-MCPT (Bayesian)} model exhibits similar predictive performance across the two subsets. The distributional results provide further insight into these patterns. On the street-test subset, stacking performs worse than \textit{Street-MCPT (Bayesian)}, whereas on the lab-test subset, it substantially improves predictive performance over the street-based model. This trade-off results in a higher overall predictive score in the pooled domain. On the lab-test subset, \textit{Street-MCPT (Bayesian)} exhibits a pronounced left tail, indicating severe prediction failures on a subset of observations, whereas \textit{Lab-MCPT (Bayesian)} shows a more concentrated distribution. The stacking model lies between the two, reducing extreme errors while maintaining reasonable predictive accuracy.

\begin{table}[H]
\centering
\caption{Predictive performance (Avg. Log Score) in Target Domain 2 (averaged over 10 data splits)}
\label{tab:domain222}
\begin{tabular}{lccc}
\hline
Model 
& Street-test 
& Lab-test 
& Pooled \\
\hline
Street-MCPT (Bayesian)
& -0.4666 
& -0.8625 
& -0.6249 \\

Lab-MCPT (Bayesian)
& -0.7363 
& -0.7420 
& -0.7386 \\ 

Stacking (PSIS-LOO) 
& -0.5152 
& -0.7602 
& \textbf{-0.6132} \\
\hline
\end{tabular}
\end{table}

\begin{figure}[H]
    \centering
    \includegraphics[width=0.6\textwidth]{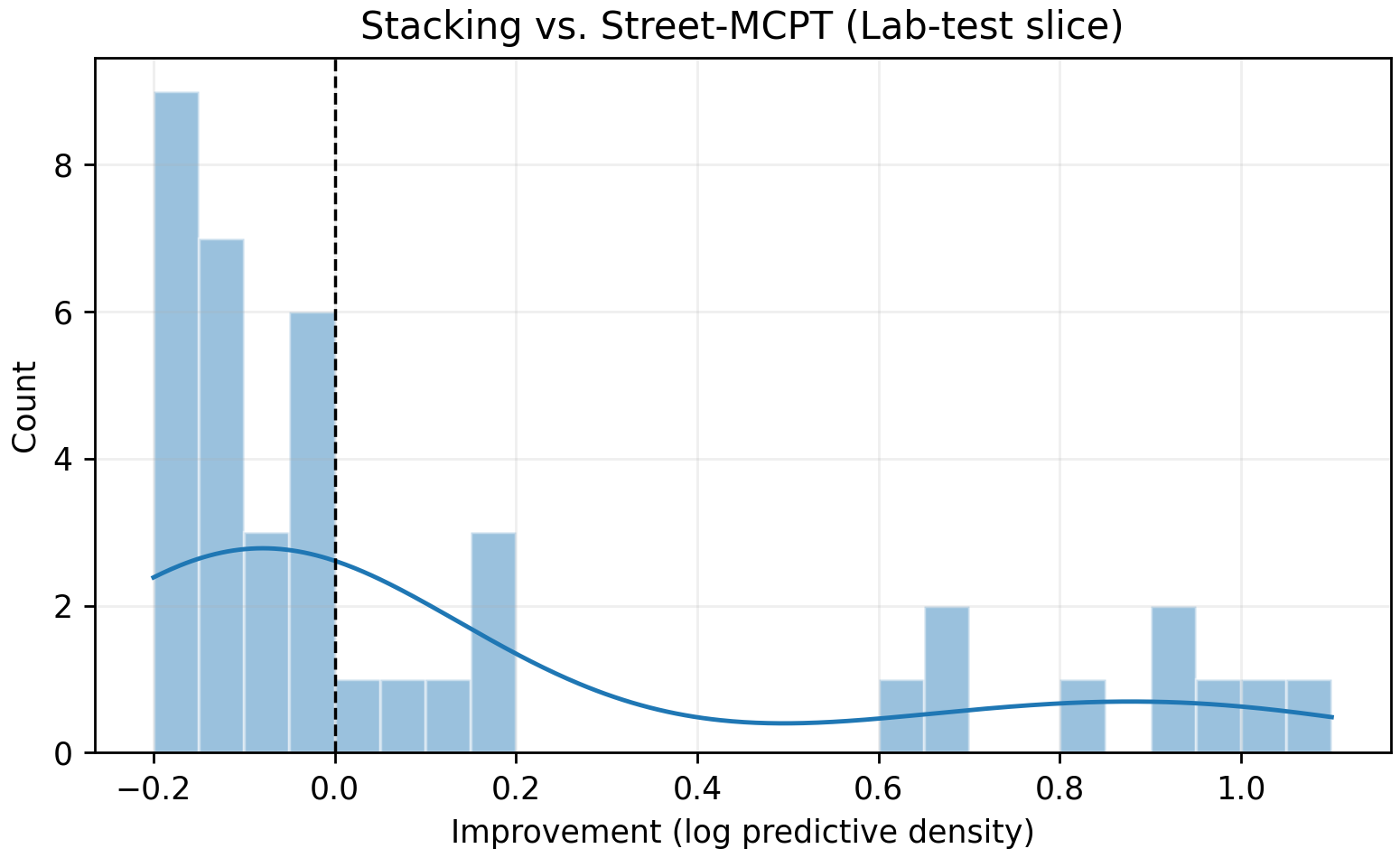}
    \caption{Distribution of pointwise improvement of stacking over Street-MCPT (Bayesian) on the lab-test slice in Target Domain 2}
    \label{fig:improvement_histogram}
\end{figure}

To further understand where the performance gains originate, Figure \ref{fig:improvement_histogram} presents the distribution of the pointwise improvement of stacking over \textit{Street-MCPT (Bayesian)} on the lab-test subset. We focus on this subset because it is where the improvement over \textit{Street-MCPT (Bayesian)} is most substantial, making it the most informative setting for understanding the gains from model fusion. This subset provides an illustrative example of cross-domain prediction. The pointwise improvement of stacking over \textit{Street-MCPT (Bayesian)} on the lab-test subset is:

\[
\Delta_i = \log p_{\text{stack}}(y_i) - \log p_{\text{Street}}(y_i).
\]

The histogram shows that stacking does not uniformly improve performance; most observations exhibit small negative differences (left tail). However, the distribution has a heavy right tail, indicating substantial gains for a subset of observations where \textit{Street-MCPT (Bayesian)} performs particularly poorly. These large positive improvements outweigh the small degradations elsewhere, leading to an overall increase in average log predictive performance.

These results suggest that the predictive gains from stacking arise primarily from mitigating severe prediction failures of \textit{Street-MCPT (Bayesian)} on a subset of laboratory observations, rather than uniformly improving all observations.

\subsection{Sensitivity to pooled target-domain composition}
\label{subsec:4.4}  
Since the pooled target domain combines observations from the laboratory and street-intercept datasets, the learned fusion weights may depend on the relative contribution of the two domains. In the main analysis, this composition is determined by the available choice tasks, resulting in approximately 40\% lab (40$\times$0.8$\times$5 = 160 choice tasks) and 60\% (152$\times$0.8$\times$2 = 244 choice tasks) street observations. To examine the sensitivity of the PSIS-LOO stacking weights to this composition, we fixed the weight-learning sample at 300 choice tasks and varied the laboratory share from 20\% to 50\%. For each composition and train-test split, participants were randomly subsampled 100 times, while the previously estimated Lab-MCPT and Street-MCPT models were used throughout. All resulting weights were evaluated on the same original pooled test set, thereby isolating the effect of the composition used for weight learning.

\begin{figure}[H]
    \centering
    \includegraphics[width=0.9\textwidth]{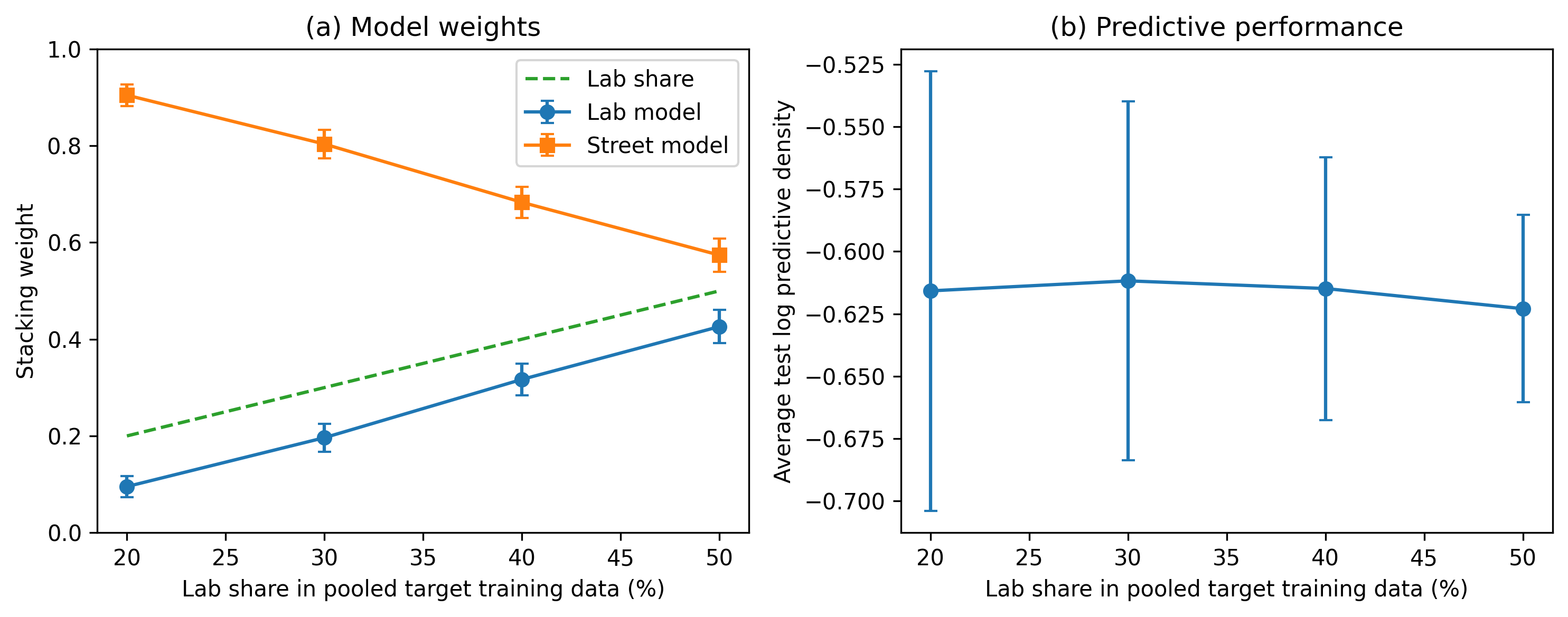}
    \caption{PSIS-LOO stacking weights and predictive performance under different compositions of the pooled target domain.}
    \label{fig:composition_sensitivity}
\end{figure}

As shown in Figure~\ref{fig:composition_sensitivity} and Table~\ref{tab:composition_sensitivity}, the stacking weights adjust systematically with the target-domain composition. The lab model weight increases from 0.0952 to 0.4261 as the laboratory share increases from 20\% to 50\%, while remaining below the corresponding lab share in all cases. This indicates that stacking responds to changes in target-domain composition without simply reproducing the relative sample proportions. Importantly, the average test log score remains largely stable across the different compositions, ranging only from $-0.6118$ to $-0.6229$. At the same time, the variance decreases from 0.306 to 0.105 as the laboratory share increases, indicating less dispersion in predictive performance across test observations. These values are also comparable to the main pooled-domain result ($-0.6132$), which uses all 404 available training tasks (244 street-intercept and 160 laboratory tasks) for weight learning, despite the smaller weight-learning sample used in the sensitivity analysis. Overall, the results suggest that while the fusion weights adapt to the assumed target-domain composition, predictive performance remains relatively robust.
\begin{table}[htbp]
\centering
\caption{Sensitivity of PSIS-LOO stacking to the composition of the pooled target domain.}
\label{tab:composition_sensitivity}
\begin{tabular}{cccccccc}
\hline
Lab share & Street share & Lab tasks & Street tasks &
$w_{\mathrm{Street}}$ & $w_{\mathrm{Lab}}$ & Test avg. & Test var. \\
\hline
20\% & 80\% & 60  & 240 & 0.9048 & 0.0952 & -0.6158 & 0.3055 \\
30\% & 70\% & 90  & 210 & 0.8036 & 0.1964 & -0.6118 & 0.2241 \\
40\% & 60\% & 120 & 180 & 0.6831 & 0.3169 & -0.6148 & 0.1525 \\
50\% & 50\% & 150 & 150 & 0.5739 & 0.4261 & -0.6229 & 0.1051 \\
\hline
\end{tabular}
\end{table}

\subsection{Logit-based individual weighting for model stacking}
\label{subsec:4.5}
To account for heterogeneity in the relative predictive performance of modality-rich and modality-lean data, we extend the standard stacking framework by introducing a logit-based formulation of model weights. While standard stacking estimates a single global weight shared across all observations, this assumption may be restrictive in our setting, where individuals are exposed to different experimental conditions.

We therefore specify individual-level weights using a binary logit model, where the street-based model is taken as the reference alternative. The resulting weights are:
\begin{equation}
w_{n,\text{lab}} = \frac{\exp(\xi_n)}{1 + \exp(\xi_n)}, \quad
w_{n,\text{street}} = 1 - w_{n,\text{lab}},
\end{equation}
where $n$ indexes individuals and $\xi_n$ is a linear function of individual-level characteristics. The resulting weights determine the relative contribution of the two predictive distributions for each individual. These weights are mapped to each observation in the panel data via $n(i)$, yielding the combined predictive distribution:
\begin{equation}
p(y_i \mid x_i) =
w_{n(i),\text{street}} \, p_{\text{street}}(y_i)
+
w_{n(i),\text{lab}} \, p_{\text{lab}}(y_i)
\end{equation}

In our implementation, the individual-specific weights are parameterised using four individual-level characteristics, namely driving frequency, daily driving distance, attitudinal preferences, and driving experience. These variables are selected to capture heterogeneity in the relative predictive reliability of the laboratory and street models across individuals. To ensure stable estimation, mild regularisation and numerical stabilisation (i.e., ridge penalty and probability clipping) are applied.

From Table \ref{tab:stacking_comparison}, logit-based stacking consistently improves predictive performance over standard stacking, with more pronounced gains in the pooled domain (Target Domain 2). In Target Domain 1, standard stacking collapses to a degenerate solution assigning full weight to the street-based model, whereas logit-based stacking introduces a non-zero contribution from the lab-based model, though the performance gain remains modest.

In contrast, the pooled setting enables both models to contribute more evenly, resulting in improvements in both predictive accuracy and stability. This indicates that modality-rich and modality-lean datasets provide complementary predictive information, which can be more effectively exploited through individual-specific weighting.

\begin{table}[htbp]
\centering
\caption{Comparison between stacking and logit-based stacking across the two target domains (average over 10 splits). }
\label{tab:stacking_comparison}
\begin{tabular}{llccccc}
\hline
Domain & Method & Avg. log score & Variance & Avg. $w_{\text{street}}$ & Avg. $w_{\text{lab}}$\\
\hline
Target Domain 1 & Stacking & -0.4666 & 0.2101  & 1.0000 & 0.0000 \\
         & Stacking-logit    & -0.4617 & 0.1531 & 0.7938 & 0.2062 \\
\hline
Target Domain 2 & Stacking & -0.6132 & 0.1502    & 0.6899 & 0.3101   \\
         & Stacking-logit    & -0.5824 & 0.1293 & 0.6018 & 0.3982 \\
\hline
\end{tabular}
\end{table}

While the above results demonstrate the predictive benefits of individual-specific weighting, an additional question is whether the weighting covariates themselves may indirectly capture attention variability. We conducted an exploratory Spearman correlation analysis between the four weighting covariates and the average attention allocation across the laboratory dataset. Specifically, for each individual, the task-level attention weights were averaged across all laboratory choice tasks to obtain one average attention measure for each attribute ($A_{RC}, A_{OC1}, A_{OC2}, A_{DR}$). As shown in Table \ref{tab:attention_covariate_correlations}, the observed associations were generally weak, with absolute correlation coefficients mostly below 0.3. Only one moderate association reached statistical significance, between driving experience and attention to driving range ($\rho=-0.316$, $p<0.05$), whereas all remaining correlations were statistically insignificant.

Overall, these results provide limited evidence that the weighting covariates systematically capture attention variability. This is consistent with the intended role of the proposed logit-based weighting mechanism, which learns individual-specific model weights to adapt the relative contribution of the two predictive models.
\begin{table}[ht]
\centering
\caption{Spearman correlations between individual characteristics and average attention allocation across laboratory choice tasks. * indicates statistical significance at the 5\% level ($p<0.05$).}
\label{tab:attention_covariate_correlations}
\begin{tabular}{lcccc}
\hline
 & $A_{RC}$ & $A_{OC1}$ & $A_{OC2}$ & $A_{DR}$ \\
\hline
Driving frequency & -0.283 & 0.050 & -0.126 & -0.048 \\
\hline
Daily driving distance & -0.174 & -0.177 & 0.019 & 0.116 \\
\hline
Attitudinal preference & 0.275 & -0.024 & 0.259 & -0.060 \\
\hline
Driving experience & -0.281 & 0.282 & -0.176 & -0.316* \\
\hline
\end{tabular}
\end{table}

\section{Conclusions}
\label{sec:conclusion}
In this paper, we investigate how to leverage modality-lean and modality-rich datasets to improve predictive performance in transport choice modelling. We apply the proposed model-fusion framework for multiple heterogeneous datasets to the two stated choice datasets collected from different experimental settings but addressing the same decision problem. Among the three methods considered, PSIS-LOO stacking consistently provides the best performance. Stacking improves prediction both within a single dataset (Target Domain 1) and for the pooled dataset (Target Domain 2). More importantly, the results also show that the definition of the target domain plays an important role. Modality-rich data do not necessarily improve predictions in out-of-domain settings when used alone, but can provide substantial gains when combined with in-domain data. Observation-level analysis further reveals that stacking improves performance by correcting severe prediction failures where the individual models perform poorly, rather than uniformly improving all observations. These findings highlight the value of combining heterogeneous datasets to improve predictive generalisation by stacking. 

The sensitivity analysis further demonstrates that the learned stacking weights depend on the composition of the pooled target domain. As the proportion of laboratory observations used for weight estimation increases, the weight assigned to the laboratory-trained model increases correspondingly. Despite these changes in model weights, predictive performance on the fixed pooled test data remains relatively stable across the compositions considered. This suggests that stacking adapts its relative reliance on the candidate models to changes in the target-domain composition while maintaining similar predictive performance in this application.

We additionally explore a logit-based extension that allows model-fusion weights to vary across individuals. The results indicate that incorporating individual-level heterogeneity can further improve predictive performance relative to standard stacking. This extension therefore provides a potential direction for developing more adaptive model-fusion strategies, although further evaluation is needed to establish its performance across other datasets and settings.

Taken together, these findings demonstrate the potential of extending prediction-level model fusion, particularly PSIS-LOO stacking, to integrate complementary information from heterogeneous datasets with respect to an explicitly defined target domain, while retaining dataset-specific model structures. In transport applications, such fused predictive distributions could provide a basis for population-level forecasting and policy analysis by translating individual choice probabilities into aggregate indicators, such as predicted market shares and elasticities under hypothetical policy scenarios. Although the empirical application focuses on modality-rich and modality-lean transport choice data, the framework may also be applicable to other prediction problems involving heterogeneous or multimodal data sources.

%limitation
Several limitations should be acknowledged. First, the empirical analysis is based on two stated-choice datasets of relatively limited size and uses MCPT as the common behavioural modelling framework. Moreover, although predictive performance is evaluated on held-out individuals, these test observations originate from the same experimental datasets used to develop the candidate models and define the target domains. The framework has therefore not yet been validated on a completely independent and previously unseen dataset. Also, this proposed model fusion framework does not require candidate models to have identical structures, but we currently only apply MCPT as the behavioural models. Future research could explore different types of models like machine learning models that can better deal with process data. Future research should examine its generalizability using larger datasets, additional experimental settings with different forms of modality heterogeneity, and alternative candidate models.

Second, the proposed framework is motivated not only by differences in available modalities and experimental settings, but also by the possibility that heterogeneous datasets represent different participant populations. While prediction-level fusion allows models estimated from such datasets to be combined without requiring identical populations, the present study does not explicitly isolate or systematically examine population heterogeneity. Future work could investigate how differences in population composition and behavioural preferences affect fusion weights and predictive performance, particularly when predictions are transferred to new populations.

Third, the attention-transfer mechanism used in this study represents one possible structural alignment strategy. Exploring more sophisticated alignment approaches, including learned attention prediction or other modality alignment techniques, could be an important direction for future research.
\newpage
\section*{Code and data availability}

The code for model estimation and supplementary analyses is publicly available at
\url{https://github.com/lanlanyan0817/Heterogeneous-dataset-fusion-via-model-stacking-.git}.
The repository provides instructions for reproducing the results
presented in this study. The original datasets are available from
\url{https://github.com/jiaxuan-ding/uncertainty}.

\section*{Acknowledgments}

Funded by the European Union (ERC, IMMERSION, 101164292). 
%Funded by the European Union (GRANT DETAILS REDACTED TO MAINTAIN ANONYMITY).
Views and opinions expressed are however those of the authors only and do not necessarily reflect those of the European Union or the European Research Council. Neither the European Union nor the granting authority can be held responsible for them.

\section*{Declaration of competing interest}
The authors declare that they have no potential conflict of interest.

\section*{Declaration of generative AI and AI-assisted technologies in the manuscript preparation process}
During the preparation of this manuscript, the authors used ChatGPT (OpenAI) to assist with language editing, improving the clarity and structure of the text, and supporting code development and documentation. All AI-assisted outputs were reviewed, verified, and revised by the authors as appropriate. The authors take full responsibility for the content of the published article.

\newpage

\appendix
\renewcommand{\thesection}{Appendix \Alph{section}}
\section{Implementation of baseline models}
\label{app:baseline_implementation}

\paragraph{Target Domain 1.}
Two baseline models are considered. The \textit{Street-MCPT} baseline is estimated using the street-intercept training data and subsequently evaluated on the held-out street test data. The \textit{Pooled-MCPT} baseline is estimated jointly using the street and lab training data under the attention-informed MCPT specification. For lab observations, observed task-level attention is used, whereas for street observations, where eye-tracking information is unavailable, attention is transferred using the scenario-level mean attention calculated from the lab training data. The resulting pooled model is then evaluated on the held-out street test data.

\paragraph{Target Domain 2.}
Three baseline models are considered. The \textit{Street-MCPT} and \textit{Lab-MCPT} baselines are estimated separately using the street and lab training data, respectively. The Street-MCPT follows the standard MCPT specification without attention information, whereas the Lab-MCPT incorporates observed task-level eye-tracking attention. The \textit{Pooled-MCPT} baseline is estimated jointly using both training datasets following the same attention-informed specification described above, using observed attention for lab observations and scenario-level transferred attention for street-intercept observations. All three models are subsequently evaluated on the pooled target domain consisting of the held-out street-intercept and lab test observations.

\section{MCPT model formulation and prior specifications}
\label{app:model_specification}

\subsection{MCPT formulation}

Both the street and laboratory models are based on the multi-attribute cumulative prospect theory (MCPT) specification of \cite{DING2025104608}. The compact utility representation used in the main text is expanded here to provide the complete model formulation.

Let $x_{ijks}$ denote the value of attribute $k$ in state $s$ for alternative $j$ in choice task $i$, where $s$ indexes the possible outcomes of an attribute under uncertainty. Following the original MCPT implementation in \cite{DING2025104608}, attribute-specific reference points, $\tilde{x}_{ik}$, are constructed using the conventional vehicle for deterministic attributes and the probability-weighted charging-cost outcomes for the uncertain attribute. The reference-dependent and attribute-scaled value is

\begin{equation}
z_{ijks}
=
\beta_{ik}
\left(
x_{ijks}-\tilde{x}_{ik}
\right).
\end{equation}

Here, $\beta_{ik}=\beta_k$ for the street model,
whereas $\beta_{ik}$ is attention-dependent for the
laboratory model, as specified in Eq. (\ref{eq:attention_beta}). The prospect value of each attribute state is given by

\begin{equation}
V_{ijks}
=
\begin{cases}
z_{ijks}^{\alpha},
& z_{ijks} \geq 0, \\[2mm]
-\lambda(-z_{ijks})^{\alpha},
& z_{ijks}<0,
\end{cases}
\end{equation}

where $0<\alpha<1$ is the diminishing-sensitivity parameter and
$\lambda>1$ is the loss-aversion parameter.

For attributes involving uncertain outcomes, objective state probabilities are transformed using the Prelec probability-weighting function,

\begin{equation}
\pi(p)
=
\exp\left[
-\left(-\ln p\right)^r
\right],
\qquad r>0,
\end{equation}

where $r$ is the decision-weight parameter. For an uncertain attribute with two possible states having probabilities $p_1$ and $p_2$, the cumulative decision weights are

\begin{equation}
W_1=\pi(p_1),
\qquad
W_2=\pi(p_1+p_2)-\pi(p_1).
\end{equation}

Since $p_1+p_2=1$ and $\pi(1)=1$, this reduces to

\begin{equation}
W_1=\pi(p_1),
\qquad
W_2=1-\pi(p_1).
\end{equation}

The deterministic utility of alternative $j$ is constructed by summing the decision-weighted prospect values across attributes and possible states,

\begin{equation}
U_{ij}
=
ASC_j
+
\sum_k\sum_s
V_{ijks}W_{ijks}.
\end{equation}

In the present application, the deterministic utilities of the conventional vehicle (CV) and electric vehicle (EV) alternatives are therefore written as

\begin{align}
U_{CV}
&=
V_{CV,RC}
+
V_{CV,OC}
+
V_{CV,DR},
\\
U_{EV}
&=
V_{EV,RC}
+
W_1 V_{EV,OC}
+
W_2 V_{EV,OC_1}
+
V_{EV,DR}
+
ASC_{EV}.
\end{align}

The EV charging cost has two possible states, $OC$ and $OC_1$, which share the same charging-cost sensitivity parameter $\beta_{OC}$. The reference points follow \cite{DING2025104608}: the CV attribute levels define the reference points for running cost and driving range, while the probability-weighted EV charging cost defines the common reference point for the two uncertain charging-cost states.

Assuming independently and identically distributed Type-I extreme-value errors, the resulting choice probability is

\begin{equation}
P(y_i=j)
=
\frac{\exp(U_{ij})}
{\sum_{\ell}\exp(U_{i\ell})}.
\end{equation}

\subsection{Model parameters and prior specifications}
\label{appendixA}

\begin{table}[htbp]
\centering
\caption{Parameters of the Bayesian MCPT models}
\label{tab:mcpt_parameters}
\begin{tabular}{lll}
\hline
Parameter & Description & Model \\
\hline
$ASC$ 
& Alternative-specific constant for EV 
& Both \\

$\alpha$ 
& Curvature parameter of the prospect value function 
& Both \\

$\lambda$ 
& Loss-aversion parameter 
& Both \\

$r$ 
& Prelec probability-weighting parameter 
& Both \\

$\beta_k$ 
& Constant attribute sensitivity 
& Street only \\

$\beta_{k,0}$ 
& Baseline component of attribute sensitivity 
& Laboratory only \\

$\gamma_k$ 
& Attention effect on attribute sensitivity 
& Laboratory only \\
\hline
\end{tabular}

\vspace{0.5em}
\begin{minipage}{0.93\textwidth}
\footnotesize
\textit{Note:} 
$k \in \{RC,OC,DR\}$. 
For the charging-cost attribute, $OC$ and $OC_1$ represent the two possible cost outcomes under uncertainty and share the same OC-level parameterisation. 
In the laboratory model, their task-specific sensitivities may differ because the corresponding attention measures enter separately in Eq.~\eqref{eq:attention_beta}.
\end{minipage}
\end{table}

\begin{table}[htbp]
\centering
\caption{Parameterisation and prior distributions used for Bayesian estimation}
\label{tab:mcpt_priors}
\begin{tabular}{llll}
\hline
Parameter (Model)
& Constraint 
& Parameterisation and prior 
& Implied 95\% interval\\
\hline

$ASC$ (Both)
& Unrestricted
& $ASC=\eta_{ASC},\quad \eta_{ASC}\sim\mathcal{N}(0,1)$
& $[-1.97,1.97]$\\

$\alpha$ (Both)
& $(0,1)$
& $\alpha=\operatorname{logit}^{-1}(\eta_{\alpha}),\quad
\eta_{\alpha}\sim\mathcal{N}(0,0.75)$
& $[0.19,0.81]$\\

$\lambda$ (Both)
& $>1$
& $\lambda=1+\exp(\eta_{\lambda}),\quad
\eta_{\lambda}\sim\mathcal{N}(0,0.3)$
& $[1.56,2.80]$\\

$r$ (Both)
& $>0$
& $r=\exp(\eta_r),\quad
\eta_r\sim\mathcal{N}(0,0.2)$
& $[0.67,1.48]$\\

$\beta_k$ (Street)
& $>0$
& $\beta_k=\exp(\eta_{\beta_k}),\quad
\eta_{\beta_k}\sim\mathcal{N}(0,0.3)$
& $[0.56,1.80]$\\

$\beta_{k,0}$ (Lab)
& $>0$
& $\beta_{k,0}=\exp(\eta_{\beta_{k,0}}),\quad
\eta_{\beta_{k,0}}\sim\mathcal{N}(0,0.3)$
& $[0.56,1.80]$\\

$\gamma_k$ (Lab)
& Unrestricted
& $\gamma_k=\eta_{\gamma_k},\quad
\eta_{\gamma_k}\sim\mathcal{N}(0,0.5)$
& $[-0.98,0.98]$\\

\hline
\end{tabular}

\vspace{0.5em}
\begin{minipage}{0.96\textwidth}
\footnotesize
\textit{Note:}
The implied 95\% intervals refer to the central 95\% prior intervals on the model-parameter scale after applying the corresponding transformations; they are not posterior credible intervals. The second argument of the normal distribution denotes the standard deviation.

\end{minipage}
\end{table}

The parameter constraints are chosen to preserve the behavioural interpretation of the MCPT formulation. 
The curvature parameter is restricted to $0<\alpha<1$ to represent diminishing sensitivity in the prospect value function, while $\lambda>1$ ensures that losses receive greater weight than equivalent gains and therefore represents loss aversion. The probability-weighting parameter is restricted to $r>0$, which yields a valid Prelec probability-weighting function.

For the street model, the attribute-sensitivity parameters are constrained to $\beta_k>0$ so that they scale the reference-dependent attribute deviations without reversing their signs and, consequently, the classification of gains and losses. The laboratory model similarly constrains the baseline components $\beta_{k,0}$ to be positive, while the final task-specific sensitivities remain positive through the softplus transformation in Eq.~\eqref{eq:attention_beta}. In contrast, the attention-effect parameters $\gamma_k$ are left unrestricted in sign, allowing deviations from the equal-attention benchmark to either increase or decrease the corresponding attribute sensitivity.

For Bayesian estimation, the constrained parameters are represented through unconstrained latent parameters and mapped to their admissible domains using inverse-logit or exponential transformations, as summarized in Table~\ref{tab:mcpt_priors}. Normal priors are specified on the unconstrained parameter scales. The street and laboratory models are estimated independently; thus, although the corresponding sensitivity parameters use the same prior specification, $\beta_k$ in the street model and $\beta_{k,0}$ in the laboratory model are separately estimated from their respective datasets.

\section{Sensitivity analysis on attention-transfer strategies}
\label{sensitivity_analysis_attention}
\begin{table}[H]
\centering
\caption{Sensitivity analysis of different attention-transfer strategies in Target Domain 2}
\label{tab:attention_sensitivity}
\begin{tabular}{lcccc}
\hline
Attention transfer strategy &
$w_{\mathrm{street}}$ &
$w_{\mathrm{lab}}$ &
Avg.\ log score &
Variance \\
\hline
Scenario-level mean (proposed) & 0.6899 & 0.3101 & -0.613218 & 0.152623 \\
Scenario-level median          & 0.6898 & 0.3102 & -0.613213 & 0.152588 \\
Global mean                    & 0.6902 & 0.3098 & -0.613223 & 0.152760 \\
Uniform attention ($0.25$)     & 0.6893 & 0.3107 & -0.613160 & 0.152476 \\
\hline
\end{tabular}
\end{table}
We conducted a sensitivity analysis to evaluate whether the proposed framework depends on the specific attention-transfer strategy. Besides the original scenario-level mean attention, we additionally considered three alternative transfer strategies: scenario-level median attention, global mean attention, and uniform attention. These alternatives vary in the extent to which they rely on scenario-specific attention information, while still providing the compatible input representation required to apply the laboratory model to the modality-lean dataset.

Across all four strategies, the predictive performance and learned stacking weights remained almost identical (Table \ref{tab:attention_sensitivity}), with the maximum difference in average log score below \(7 \times 10^{-5}\). This suggests that the proposed model fusion framework is robust to the specific attention-transfer strategy.  
%One possible explanation is that the attention transfer is only used within the lab model. Since the final prediction is obtained by stacking the lab and street models, rather than directly using the transferred attention, its effect is naturally diluted during the model fusion process.

Importantly, the purpose of the attention-transfer mechanism is not to recover the true (unobserved) attention process, but to resolve the structural mismatch between modality-rich and modality-lean models by providing a compatible input representation. The sensitivity analysis therefore suggests that the final model fusion results are not materially affected by the choice of attention-transfer strategy among those examined.

\newpage
\bibliographystyle{apalike}
\bibliography{bibliography.bib}

@article{DING2025104608,
title = {Preferences for electric vehicles under uncertain charging prices: An eye-tracking study},
journal = {Transportation Research Part D: Transport and Environment},
volume = {140},
pages = {104608},
year = {2025},
issn = {1361-9209},
doi = {10.1016/j.trd.2025.104608},
url = {https://www.sciencedirect.com/science/article/pii/S1361920925000185},
author = {Jiaxuan Ding and Jiangbo (Gabe) Yu and Prateek Bansal},
  note = {DOI: \url{https://doi.org/10.1016/j.trd.2025.104608}}
}

@article{yao2018using,
  title={Using stacking to average Bayesian predictive distributions (with discussion)},
  author={Yao, Yuling and Vehtari, Aki and Simpson, Daniel and Gelman, Andrew},
  year={2018},
  journal={Bayesian Analysis},
  volume={13},
  number={3},
  pages={917--1007},
  doi={10.1214/17-BA1091},
  note = {DOI: \url{https://doi.org/10.1214/17-BA1091}}
}

@techreport{burris2023examining,
  title={Examining the Impact of Stress and Fear of {COVID-19} on Transit Travel},
  author={Burris, Mark and Brown, Alexander and Gupta, Hardik and Wang, Jasper and Jha, Kartikeya and Walk, Michael and Figueroa-Medina, Alberto M. and Rodr{\'\i}guez-Rom{\'a}n, Daniel and del Valle-Gonz{\'a}lez, Carlos and Nieves-D{\'\i}az, Luis and others},
  institution={National Institute for Congestion Reduction, University of South Florida},
  type={Final Report},
  year={2023},
  doi={10.5038/CUTR-NICR-Y2-1-6},
  note = {DOI: \url{https://doi.org/10.5038/CUTR-NICR-Y2-1-6}}
}

@inproceedings{lee2022remotelab,
  title={{RemoteLab}: A {VR} remote study toolkit},
  author={Lee, Jaewook and Natarrajan, Raahul and Rodriguez, Sebastian S and Panda, Payod and Ofek, Eyal},
  booktitle={Proceedings of the 35th Annual ACM Symposium on User Interface Software and Technology},
  pages={1--9},
  year={2022},
  doi={10.1145/3526113.3545679},
  note = {DOI: \url{https://doi.org/10.1145/3526113.3545679}}
}

@article{mottelson2021self,
  title={A self-administered virtual reality intervention increases {COVID-19} vaccination intention},
  author={Mottelson, Aske and Vandeweerdt, Clara and Atchapero, Michael and Luong, Tiffany and Holz, Christian and B{\"o}hm, Robert and Makransky, Guido},
  journal={Vaccine},
  volume={39},
  number={46},
  pages={6746--6753},
  year={2021},
  publisher={Elsevier},
  doi={10.1016/j.vaccine.2021.10.004},
  note = {DOI: \url{https://doi.org/10.1016/j.vaccine.2021.10.004}}
}

@article{ton2020latent,
  title={Latent classes of daily mobility patterns: the relationship with attitudes towards modes},
  author={Ton, Danique and Zomer, Lara-Britt and Schneider, Florian and Hoogendoorn-Lanser, Sascha and Duives, Dorine and Cats, Oded and Hoogendoorn, Serge},
  journal={Transportation},
  volume={47},
  number={4},
  pages={1843--1866},
  year={2020},
  publisher={Springer},
  doi={10.1007/s11116-019-09975-9},
  note = {DOI: \url{https://doi.org/10.1007/s11116-019-09975-9}}
}

@article{bogacz2021modelling,
  title={Modelling risk perception using a dynamic hybrid choice model and brain-imaging data: an application to virtual reality cycling},
  author={Bogacz, Martyna and Hess, Stephane and Calastri, Chiara and Choudhury, Charisma F and Mushtaq, Faisal and Awais, Muhammad and Nazemi, Mohsen and van Eggermond, Michael AB and Erath, Alexander},
  journal={Transportation research part C: emerging technologies},
  volume={133},
  pages={103435},
  year={2021},
  publisher={Elsevier},
  doi={10.1016/j.trc.2021.103435},
  note = {DOI: \url{https://doi.org/10.1016/j.trc.2021.103435}}
}

@article{argota2024virtual,
  title={A virtual reality experiment to study pedestrian perception of future street scenarios},
  author={Argota S{\'a}nchez-Vaquerizo, Javier and Hausladen, Carina I and Mahajan, Sachit and Matter, Marc and Siebenmann, Michael and van Eggermond, Michael AB and Helbing, Dirk},
  journal={Scientific reports},
  volume={14},
  number={1},
  pages={4571},
  year={2024},
  publisher={Nature Publishing Group UK London},
  doi={10.1038/s41598-024-55073-x},
  note = {DOI: \url{https://doi.org/10.1038/s41598-024-55073-x}}
}

@article{derakhshan2024situated,
  title={A situated inspection of autonomous vehicle acceptance--A population study in virtual reality},
  author={Derakhshan, Shadi and Nosrat Nezami, Farbod and W{\"a}chter, Maximilian Alexander and Stephan, Achim and Pipa, Gordon and K{\"o}nig, Peter},
  journal={International Journal of Human--Computer Interaction},
  year={2024},
  publisher={Taylor \& Francis},
  volume={41},
  number={9},
  pages={5225--5244},
  doi={10.1080/10447318.2024.2358577},
  note = {DOI: \url{https://doi.org/10.1080/10447318.2024.2358577}}
}

@article{zhang2025mc,
  title={{MC-HSTA}: A multi-source cross-domain hybrid spatio-temporal attention network for traffic flow prediction},
  author={Zhang, Zhirong and Cheng, Qixiu and Li, Zhanyi and Fang, Yong and Li, Ying},
  journal={Neural Networks},
  pages={108358},
  year={2025},
  publisher={Elsevier},
  volume={196},
  doi={10.1016/j.neunet.2025.108358},
  note = {DOI: \url{https://doi.org/10.1016/j.neunet.2025.108358}}
}

@article{rahate2022multimodal,
  title={Multimodal co-learning: Challenges, applications with datasets, recent advances and future directions},
  author={Rahate, Anil and Walambe, Rahee and Ramanna, Sheela and Kotecha, Ketan},
  journal={Information Fusion},
  volume={81},
  pages={203--239},
  year={2022},
  publisher={Elsevier},
  doi={10.1016/j.inffus.2021.12.003},
  note = {DOI: \url{https://doi.org/10.1016/j.inffus.2021.12.003}}
}

@article{ding2015missing,
  title={Missing modality transfer learning via latent low-rank constraint},
  author={Ding, Zhengming and Shao, Ming and Fu, Yun},
  journal={IEEE transactions on Image Processing},
  volume={24},
  number={11},
  pages={4322--4334},
  year={2015},
  publisher={IEEE},
  doi={10.1109/TIP.2015.2462023},
  note = {DOI: \url{https://doi.org/10.1109/TIP.2015.2462023}}
}

@inproceedings{liang2021cross,
  title={Cross-modal generalization: Learning in low resource modalities via meta-alignment},
  author={Liang, Paul Pu and Wu, Peter and Ziyin, Liu and Morency, Louis-Philippe and Salakhutdinov, Ruslan},
  booktitle={Proceedings of the 29th ACM International Conference on Multimedia},
  pages={2680--2689},
  year={2021},
  doi={10.1145/3474085.3475247},
  note = {DOI: \url{https://doi.org/10.1145/3474085.3475247}}
}

@article{athanasiadis2020audio,
  title={Audio--visual domain adaptation using conditional semi-supervised generative adversarial networks},
  author={Athanasiadis, Christos and Hortal, Enrique and Asteriadis, Stylianos},
  journal={Neurocomputing},
  volume={397},
  pages={331--344},
  year={2020},
  publisher={Elsevier},
  doi={10.1016/j.neucom.2019.09.106},
  note = {DOI: \url{https://doi.org/10.1016/j.neucom.2019.09.106}}
}

@inproceedings{yang2017cross,
  title={Cross-domain and cross-modality transfer learning for multi-domain and multi-modality event detection},
  author={Yang, Zhenguo and Cheng, Min and Li, Qing and Li, Yukun and Lin, Zehang and Liu, Wenyin},
  booktitle={International Conference on Web Information Systems Engineering},
  pages={516--523},
  year={2017},
  organization={Springer},
  doi={10.1007/978-3-319-68783-4_35},
  note = {DOI: \url{https://doi.org/10.1007/978-3-319-68783-4_35}}
}

@article{shao2025cross,
  title={Cross-city traffic prediction based on deep domain adaptive transfer learning},
  author={Shao, Junqi and Li, Shen and Zhang, Ke and Wang, Anyou and Li, Meng},
  journal={Transportation Research Part C: Emerging Technologies},
  volume={176},
  pages={105152},
  year={2025},
  publisher={Elsevier},
  doi={10.1016/j.trc.2025.105152},
  note = {DOI: \url{https://doi.org/10.1016/j.trc.2025.105152}}
}

@article{sun2018analyzing,
  title={Analyzing cross-domain transportation big data of New York City with semi-supervised and active learning},
  author={Sun, Huiyu and McIntosh, Suzanne},
  journal={Computers, Materials \& Continua},
  volume={57},
  number={1},
  pages={1--9},
  year={2018},
  doi={10.32604/cmc.2018.03684},
  note = {DOI: \url{https://doi.org/10.32604/cmc.2018.03684}}
}

@article{hess2026combine,
  title={Combine and conquer: Model averaging for out-of-distribution forecasting},
  author={Hess, Stephane and van Cranenburgh, Sander},
  journal={Transportation Research Part C: Emerging Technologies},
  volume={192},
  pages={105846},
  year={2026},
  publisher={Elsevier},
  doi={10.1016/j.trc.2026.105846},
  note = {DOI: \url{https://doi.org/10.1016/j.trc.2026.105846}}
}

@article{avineri2004cumulative,
  title={A cumulative prospect theory approach to passengers behavior modeling: Waiting time paradox revisited},
  author={Avineri, Erel},
  journal={Journal of Intelligent Transportation Systems: Technology, Planning, and Operations},
  volume={8},
  number={4},
  pages={195--204},
  year={2004},
  publisher={Taylor \& Francis},
  doi={10.1080/15472450490523856},
  note = {DOI: \url{https://doi.org/10.1080/15472450490523856}}
}

@article{de2011expected,
  title={Expected utility theory, prospect theory, and regret theory compared for prediction of route choice behavior},
  author={de Moraes Ramos, Giselle and Daamen, Winnie and Hoogendoorn, Serge},
  journal={Transportation Research Record},
  volume={2230},
  number={1},
  pages={19--28},
  year={2011},
  publisher={SAGE Publications Sage CA: Los Angeles, CA},
  doi={10.3141/2230-03},
  note = {DOI: \url{https://doi.org/10.3141/2230-03}}
}

@article{tversky1992advances,
  title={Advances in prospect theory: Cumulative representation of uncertainty},
  author={Tversky, Amos and Kahneman, Daniel},
  journal={Journal of Risk and uncertainty},
  volume={5},
  number={4},
  pages={297--323},
  year={1992},
  publisher={Springer},
  doi={10.1007/BF00122574},
  note = {DOI: \url{https://doi.org/10.1007/BF00122574}}
}

@article{xu2011decision,
  title={A decision-making rule for modeling travelers’ route choice behavior based on cumulative prospect theory},
  author={Xu, Hongli and Zhou, Jing and Xu, Wei},
  journal={Transportation Research Part C: Emerging Technologies},
  volume={19},
  number={2},
  pages={218--228},
  year={2011},
  publisher={Elsevier},
  doi={10.1016/j.trc.2010.05.009},
  note = {DOI: \url{https://doi.org/10.1016/j.trc.2010.05.009}}
}

@article{ben1990estimation,
  title={Estimation of switching models from revealed preferences and stated intentions},
  author={Ben-Akiva, Moshe and Morikawa, Takayuki},
  journal={Transportation Research Part A: General},
  volume={24},
  number={6},
  pages={485--495},
  year={1990},
  publisher={Elsevier},
  doi={10.1016/0191-2607(90)90037-7},
  note = {DOI: \url{https://doi.org/10.1016/0191-2607(90)90037-7}}
}

@article{TABASI2023100632,
title = {Empowering revealed preference survey with a supplementary stated preference survey: demonstration of willingness-to-pay estimation within a mode choice case},
journal = {Travel Behaviour and Society},
volume = {33},
pages = {100632},
year = {2023},
issn = {2214-367X},
doi = {10.1016/j.tbs.2023.100632},
url = {https://www.sciencedirect.com/science/article/pii/S2214367X23000832},
author = {Maliheh Tabasi and Alireza Raei and Tim Hillel and Rico Krueger and Taha {Hossein Rashidi}},
  note = {DOI: \url{https://doi.org/10.1016/j.tbs.2023.100632}}
}

\end{document}